\documentclass[natbib,smallextended]{svjour3}       
\usepackage[affil-it]{authblk}
\usepackage{graphicx}
\usepackage[space]{grffile}
\usepackage{latexsym}
\usepackage{textcomp}
\usepackage{longtable}
\usepackage{multirow,booktabs}
\usepackage{amsfonts,amsmath,amssymb}
\usepackage{natbib}
\usepackage{url}
\usepackage{hyperref}
\usepackage{abbreviations}
\newif\iflatexml\latexmlfalse
\DeclareGraphicsExtensions{.pdf,.PDF,.png,.PNG,.jpg,.JPG,.jpeg,.JPEG}

\usepackage[utf8]{inputenc}
\usepackage[ngerman,english]{babel}

\begin{document}

\title{Modelling Jets, Tori and Flares in Pulsar Wind Nebulae}

\author{Oliver Porth \and Rolf Buehler \and Barbara Olmi \and Serguei Komissarov \and Astrid Lamberts \and Elena Amato \and Yajie Yuan \and Alexander Rudy}
\institute{
  Oliver Porth \at 
  Institute for Theoretical Physics, Max-von-Laue Str. 1, D-60438 Frankfurt am Main, Germany. \email{porth@itp.uni-frankfurt.de}
  \and
  Rolf Buehler \at 
  DESY, Platanenallee 6, D-15738 Zeuthen, Germany
  \and
  Barbara Olmi \at 
  Dipartimento di Fisica ed Astronomia, Universit\`a degli Studi di Firenze, Via G. Sansone 1, 50019 Sesto F.no (Firenze), Italy \\
  INFN- Sezione di Firenze, Via G. Sansone 1, 50019 Sesto F.no (Firenze), Italy \\
  INAF Osservatorio Astrofisico di Arcetri, Largo E. Fermi 5, 50125 Firenze, Italy
  \and
  Serguei Komissarov \at 
  School of Mathematics, University of Leeds, Leeds, LS29JT, UK
  \and
  Astrid Lamberts \at
  Theoretical Astrophysics, California Institute of Technology, Pasadena, CA 91125, USA
  \and
  Elena Amato \at 
  INAF Osservatorio Astrofisico di Arcetri, Largo E. Fermi 5, 50125 Firenze, Italy\\
  Dipartimento di Fisica ed Astronomia, Universit\`a degli Studi di Firenze, Via G. Sansone 1, 50019 Sesto F.no (Firenze), Italy
  \and
  Yajie Yuan \at 
  Lyman Spitzer, Jr. Postdoctoral Fellow, Department of Astrophysical Sciences, 4 Ivy  Lane Princeton University, Princeton, NJ 08544, USA \\
  KIPAC, SLAC and Stanford University, Stanford CA 94305 USA
  \and
  Alexander Rudy \at 
  Department of Astronomy \& Astrophysics, University of California, Santa Cruz, CA 95064, USA
}

\authorrunning{Porth, Buehler, Olmi, Komissarov, Lamberts, Amato, Yuan, Rudy}
\titlerunning{Modelling of PWN}

\date{\today}

\maketitle

\section{Introduction}
At the end of their life, massive stars leave behind some of the most powerful high energy sources in the sky: neutron stars or black holes. The environment around these fascinating objects allows
us to test conditions which often can not be replicated on Earth. Among these systems, pulsar wind nebulae (PWN) are of particular interest. In them, the relativistic plasma wind ejected by the rotating neutron star interacts with the ambient medium. The latter is in most cases composed of the ejecta of the original star explosion. This interaction leads to bright synchrotron and inverse Compton (IC) emission of the wind electrons, producing some of the brightest sources in the sky (electrons and positrons are referred to together as electrons throughout this text).

Modeling the emission of PWN reveals that these systems are able to accelerate particles to very high energies of up to several PeV. The ultimate energy source of this acceleration is rotation of the neutron star which induces a strong electric potential. As neutron stars are highly magnetized and conducting, they act as a unipolar inductor, creating an electric potential between poles and equator. This electromagnetic energy is carried away from the pulsar magnetosphere with the pulsar wind. Where and how it is transformed into kinetic particle energy is a matter of intense research.

\begin{figure}[h!]
\begin{center}
\includegraphics[width=1\columnwidth]{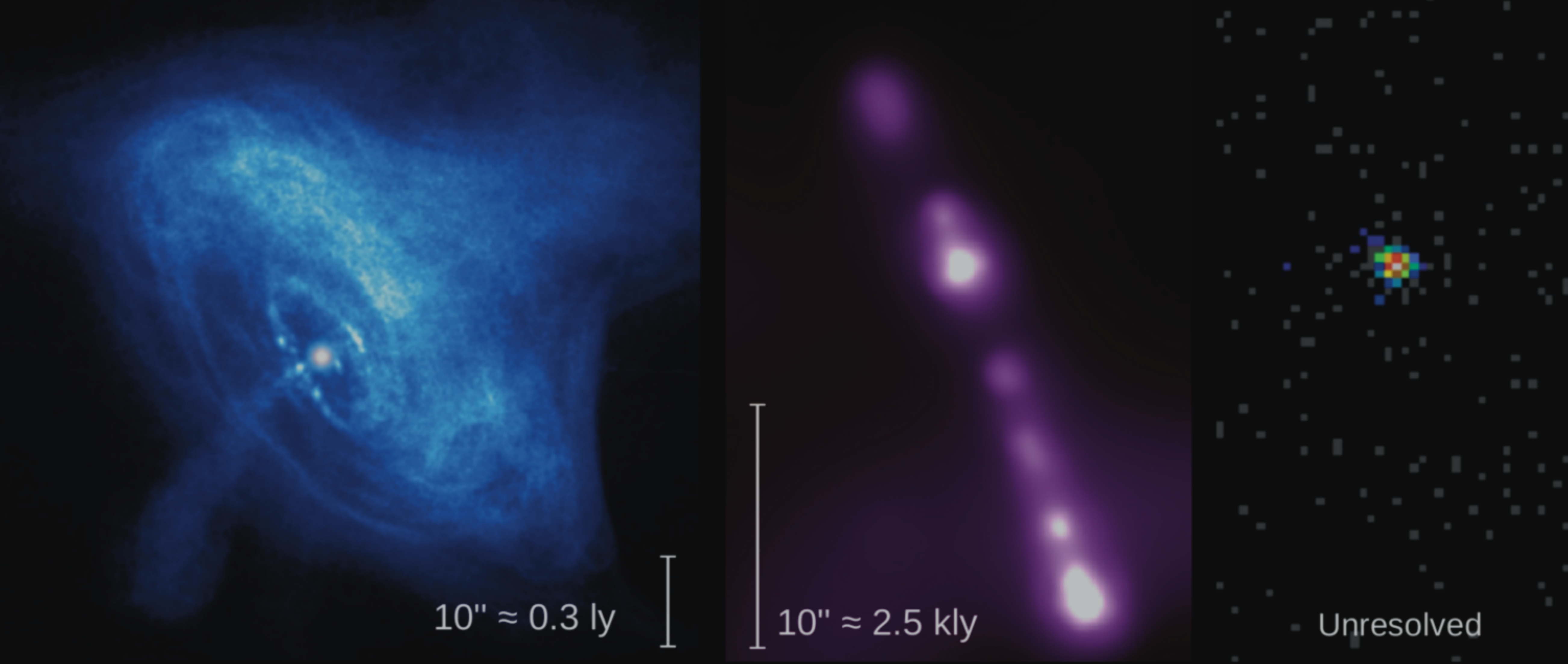}
\caption{{\label{fig:rescomp}
Images of sources of relativistic plasma taken by the Chandra X-ray Observatory (CXO): the right panel shows the Crab nebula, the middle panel the AGN M87 and the right panel GRB 991216. Each panel shows the approximate spatial scale which can be resolved within the Chandra angular resolution CXO of $\approx1$ arc sec. Image credits: NASA/CXC/ASU/J. Hester et al./E.Perlmanet al./L.Piro et al..
}}
\end{center}
\end{figure}

In comparison to other sources of relativistic plasma, PWN can be resolved in great detail, as shown in Figure \ref{fig:rescomp}. Scales close to the gyroradius of the highest energy electrons in this system can be resolved by current radio, optical and X-ray telescopes. Particularly in two systems, the Crab and Vela PWN, the plasma flow can be resolved down to scales of $\approx 10$ light days. These systems provide perfect test beds to study the behaviour of relativistic magnetized plasma, which is thought to be present in other high energy sources as gamma-ray bursts or active galactic nuclei \citep{Berger_2014,Massaro_2015}. The discovery of gamma-ray flares from the Crab PWN revealed very efficient and rapid acceleration of particles in this system. Strong and rapid gamma-ray flares are also observed in GRBs and AGN (see e.g. \citet{Ackermann_2013} and \citet{Aharonian_2009}). It is likely, that common mechanisms, magnetic reconnection or shock acceleration are responsible for the acceleration in all of these systems (see \citet{Kagan_2015} and \citet{Sironi_2015} for recent reviews).

In contrast to the inner region of the Vela PWN, the Crab PWN has been detected across the electromagnetic spectrum. The plasma motion is resolved in space and time with unmatched resolution in this system (see \citet{hester2008} and \citet{BuehlerBlandford2013a} for reviews). It has therefore been the target of most theoretical studies of PWN. In particular, relativistic magnetohydrodynamic (RMHD) simulations have provided deep insights into the plasma behaviour over the past decades (see e.g.  \citet{ssk-lyub-04,delzanna-06,PorthKomissarov2013}). The qualitative behavior of the global plasma flow in these systems is thought to be understood to date. The energy of the wind emitted by the pulsar is concentrated towards the equator. The wind is thought to be highly magnetized and cold (meaning that its thermal energy is much smaller than its magnetic and bulk kinetic energy). Over time, this wind has blown a bubble into the ejecta of the original star explosion. A reverse shock emerges when the pulsar wind first interacts with the surrounding medium, creating an oblate termination surface (see {Figure \ref{fig:rmhdflow}} in Section \ref{sec:rmhd}). Downstream of this termination surface the plasma flows into the equatorial region forming a torus and towards the poles. The latter flow is a result of the magnetic hoop stress and forms the jet observed perpendicular to the torus.

In this article, we will review the status of our current understanding of the plasma flow in PWN and the particle acceleration happening within it. The article is structured as follows: first, we will introduce the RMHD models of PWN in Section \ref{sec:rmhd}. We then proceed to discuss the particle acceleration in PWN. In Section \ref{sec:wisps} we discuss what can be learned about the particle acceleration from the dynamical structures called ``wisps'' observed in the Crab nebula. Then we proceed to discuss the gamma-ray flares observed in this system in Section \ref{sec:flares}. In this context, we will also discuss recent observational and theoretical results of the inner knot of the Crab nebula in Section \ref{sec:knot}, which had been proposed as the emission site of the flares. In Section \ref{sec:binaries} we will extend the discussion to binary systems, in which the pulsar wind interacts with the stellar wind from a companion star. The article concludes with a discussion of solved and unsolved problems of our understanding of PWN in Section \ref{sec:discussion}.

\section{RMHD Models of PWN}
\label{sec:rmhd}
\subsection{Simulating the plasma flow}
PWN are customarily modelled under assumption of ideal magnetohydrodynamics (MHD).  Although the validity of the fluid approximation in PWN is not without question\footnote{For example, on the scale of the gyroradius for X-ray emitting particles, significant wisp-like substructure is observed which prompts for a more accurate description taking into account finite gyroradius effects (MHD+PIC).}, by upholding basic conservation laws, a first insight into the physics can be attained.  Let us start with a brief review of the neutron star magnetosphere and then make our way out to the termination shock and PWN contact discontinuity. 

\subsection{Magnetosphere}

MHD can be applied if sufficient plasma is available to screen the electric field: $\mathbf{E\cdot B=0}$.  Seminal works of \citet{deutsch1955} and \citet{Goldreich_1969} indicate that this is a good working assumption as an electromagnetic cascade develops that fills the ``gaps'' of unscreened electric fields with freshly created pair-plasma \citep{Sturrock1971}.  
The number density of pairs divided by the minimal density required to screen the electric fields in the co-moving frame is known as pulsar multiplicity $\kappa$ \citep{Goldreich_1969}.  
Due to the supply of conducting magnetospheric plasma, the early vacuum oblique dipole models \citep[e.g.][]{pacini1967,Ostriker:1969} are not applicable, rather can the pulsar be thought of as a \emph{unipolar inductor} whose rotation drives a current system.  Torque from the $\mathbf{j\times B}$ Lorentz force of the current flowing parallel to  the neutron star surface serves to spin down the pulsar and extracts energy carried away in the form of Poynting flux.  
In the limit that EM contributions are stronger than pressure, inertial and gravitational forces, plasma can be considered force-free, that is the EM forces balance exactly: 
\begin{align}
\frac{1}{c}\, \mathbf{j\times B}+\rho_e \mathbf{E}=0\ . \label{eq:force-free}
\end{align}
In this regime, an important exact solution was obtained by \cite{1973ApJ...180L.133M}.  In Michel's \emph{monopolar} solution, field lines are confined to cones in the $\varpi z$-planes and describe perfect Archimedean spirals in all $\varpi\phi$-planes ($\varpi$ is the cylindrical radius).  Moving away from the source, the magnetic field is thus progressively winding up with $B_\phi=B_0\varpi\Omega/c\ (r/r_0)^{-2}$, $B_r=B_0(r/r_0)^{-2}$ ($\Omega$ is the angular velocity of the central object).  The induced electric field is simply $E_\theta = \varpi \Omega B_r/c$ and we obtain a Poynting flux 
\begin{align}
\mathcal{S}_r=\frac{c}{4\pi} \mathbf{E\times B} = \frac{\Omega^2 B_0^2 r_0^4 \sin^2\theta  }{4\pi r^2}\, . \label{eq:michelpower}
\end{align}
The important point to note here is that the energy flux is highly anisotropic $\propto \sin^2\theta$, thus more energy is ejected in the equatorial direction.  
After long struggle, a solution to the at first glance unimposing equation (\ref{eq:force-free}) for more realistic \emph{dipolar} stellar field was obtained numerically by \cite{1999ApJ...511..351C}.  
Most notably, the dipole field rips open at the light cylinder radius $\varpi_{\rm LC} = c/\Omega$ and a radial wind streams out similar to Michel's solution i.e. with $\mathcal{S}_r\propto \sin^2\theta$.  This general result for the aligned rotator has been confirmed and improved on by many groups \citep{2004MNRAS.349..213G,2005PhRvL..94b1101G,2006MNRAS.368L..30M,ParfreyBeloborodov2012,RuizPaschalidis2014}.  
As the polarity of the dipole field reverses across hemispheres, so does the wound-up toroidal field, giving rise to an equatorial current-sheet.  The ensuing dissipation violates the ideal MHD condition $\mathbf{E\cdot B=0}$ and also force-freeness must break down locally as magnetic dominance cannot be maintained.  
It was noted already by \cite{coroniti1990} that the current sheet can play an important role in the dynamics of the wind as a whole.  In case the magnetic axis is mis-aligned with the rotation axis (oblique rotator), the current-sheet assumes a wavy or \emph{striped}-shape.  Dissipation, dynamics and particle acceleration of the striped wind is subject to active research and worthy of a review of its own \citep[see ][]{arons2012}.  We shall return to this issue below.  

The first force-free model of an \emph{oblique} dipole magnetosphere was presented by \cite{spitkovsky2006} and has been confirmed by several groups and methodologies \citep{kalapotharakos2012,2016MNRAS.455.3779P,2016MNRAS.457.3384T}.  Its salient features are the complex geometry of the  striped wind \citep[as predicted by][]{Michel1971} and the dependence of the spin-down power on the obliquity angle $\alpha$
\begin{align}
L = k_1 \frac{\mu^2 \Omega^4}{c^3}(1+k_2\sin^2 \alpha) \label{eq:oblique-spindown}
\end{align}
Here $\mu=B_p r^{\star3}/2$ is the magnetic moment of the star ($B_p$ measured at the poles) and $k_1=1 \pm 0.05$, $k_2 =1\pm 0.1$ were obtained by fitting to the numerical simulations.  Following Eq. (\ref{eq:oblique-spindown}), oblique pulsars emit up to twice as much wind-power as aligned rotators.  Concerning the distribution of wind power in the orthogonal case, one obtains 
\begin{align}
\langle \mathcal{S}_r\rangle_\phi^{90^\circ} \approx \frac{\Omega^2 B_0^2 r_0^4 \sin^4\theta}{8\pi r^2} \label{eq:obliquesr}
\end{align}
where the extra factor of $\sin^2\theta$ comes from bunching of radial field in the equatorial regions.  Semi-analytic formulae for intermediate cases were presented by \cite{2016MNRAS.457.3384T}.  

Recent global particle-in-cell (PIC) simulations now confirm that given sufficient plasma supply, the magnetosphere adopts a near force free configuration consistent with the fluid models \citep{PhilippovSpitkovsky2014,ChenBeloborodov2014,Belyaev2015,CeruttiPhilippov2015} showing that the above predictions are likely robust.  

\subsection{Wind zone}

To estimate the plasma-parameters in the PWN, it is instructive to first study a 1D description of the wind flow.  
Motivated by the success of Michel's solution, we assume that the wind predominantly streams out in spherical r-direction, $\mathbf{B}=B_r\mathbf{\hat{e}}^r+B_\phi\mathbf{\hat{e}}^\phi$, $\mathbf{v}=v_r\mathbf{\hat{e}}^r+v_\phi\mathbf{\hat{e}}^\phi$, the relevant equations for the stationary flow are
\begin{align}
  \partial_r(r^2\Gamma \rho v_r) &= 0 \label{eq:continuity} \\
  \partial_r(r^3(\omega v_\phi v_r \Gamma^2-B_\phi B_r)) &= 0 \label{eq:mphi}\\ 
  \partial_r(r^2 (\omega \Gamma^2 v_r + \mathcal{S}_r)) &= 0 \label{eq:energy} \\
  \partial_r(r E_\theta) &= 0 \label{eq:induction} \\
  \partial_r(r^2 B_r) &= 0 \label{eq:divb}\ .
\end{align}
which describe conservation of mass, angular-momentum, energy, Faraday's law of induction and the $\mathbf{\nabla\cdot B}=0$ constraint.  
Adopting an ideal equation of state with adiabatic index $\hat{\gamma}$ we write the enthalpy-density $\omega=\rho c^2+\hat{\gamma}/(\hat{\gamma}-1) p $.  In the above equations, the wind Poynting flux is given by $\mathbf{\mathcal{S}} = \mathbf{E \times B}$ (hence setting $4\pi=c=1$).

Let's derive some useful relations from the system (\ref{eq:continuity}) - (\ref{eq:divb}).  
From the $\mathbf{\nabla\cdot B}=0$ constraint (\ref{eq:divb}), we immediately see that the radial field must decay as $B_r\propto r^{-2}$.    From the induction-law (\ref{eq:induction}) together with the ideal MHD electric field $E_\theta=B_\phi v_r-B_r v_\phi$ and $B_r\propto r^{-2}$, we obtain a first conservation law:
\begin{align}
  r\Omega \equiv v_\phi - v_r \frac{B_\phi}{B_r} \label{eq:romega}
\end{align}
It means that the rotation of the central object  $\Omega$  is conserved as ``angular velocity of the field lines''.  
Equation (\ref{eq:romega}) can serve to visualise two previously mentioned aspects of the pulsar magnetosphere:  1. In Michel's solution, were the field-lines rigid radial wires sticking out of the spinning source, beyond the light-cylinder plasma would be forced to rotate faster than the speed of light.  This is circumvented by induction of a toroidal component $B_\phi$ and winding up of the field.  2. In the closed dipolar magnetosphere, as $v_r=0$ on the equator, again we would also obtain $v_\phi>c$, thus field lines crossing the light-cylinder must be open.  
Furthermore, from (\ref{eq:romega}) we can estimate that at the light-cylinder $B_\phi\simeq B_r$.  

Dividing the toroidal momentum-flux (\ref{eq:mphi}) by the particle rest-mass energy flux $r^2\Gamma \rho v_r c^2$ (\ref{eq:continuity}), we obtain the conserved angular momentum flux
\begin{align}
  l \equiv \frac{\omega\Gamma r v_\phi}{\rho c^2}-\frac{r B_\phi}{k c^2}
\end{align}
with the ratio of matter to magnetic flux $k\equiv\rho v_r \Gamma / B_r$ that is also constant along a streamline.  For an accelerating wind close to the speed of light ($v_r\simeq c$) we now see that $B_\phi/B_r\simeq r/r_{LC}$ and the wind becomes dominated by the toroidal magnetic field.  Similarly, the wind rotation follows $v_\phi\propto r^{-1}$ and far away from the light-cylinder, the flow is well described by a purely radial velocity $v_r$ and an entirely toroidal magnetic field $B_\phi$.

Dividing the energy flux (\ref{eq:energy}) by the rest-mass energy flux, the conserved quantity
\begin{align}
  \mu \equiv \frac{\omega \Gamma}{\rho c^2} + \frac{\mathcal{S}_r}{\Gamma \rho v_r c^2} = \Gamma (\omega/\rho c^2 + \sigma)
\end{align}
is recovered.  It is clear that the Lorentz factor of the wind cannot exceed the value of $\mu$.  In the latter equation we have introduced the magnetization or $\sigma$-parameter:
\begin{align}
  \sigma\equiv \mathcal{S}_r/(\Gamma^2\rho v_r c^2)
\end{align}
which compares the Poynting flux with the kinetic energy of the wind.  In the cold limit $\omega\to \rho c^2$, we have $\mu=\Gamma(1+\sigma)$ which shows that accelerating the wind goes hand-in-hand with decreasing its magnetization.

As we shall see later, the value for $\sigma$ at the TS as inferred from 1D and 2D models is $\sigma\sim10^{-3}-10^{-2}$.  Quite in contrast, models for the pulsar magnetosphere predict highly magnetised plasma with $\sigma\sim10^3$ \citep[e.g.][and references therein]{arons2012}.  
The conversion of magnetic energy to kinetic energy in both confined and unconfined winds has been a subject of intensive research.  
Although claims have been made that the ideal MHD acceleration can provide the required energy conversion, \cite[e.g.][]{vlahakis2004}, it is now widely accepted that relativistic MHD flows are very inefficient accelerators \citep[e.g.][]{2009MNRAS.394.1182K,2009ApJ...699.1789T,lyubarsky2009,Lyubarsky2010} and can achieve $\sigma\approx1$ at best.  
The discrepancy between $\sigma$ obtained via MHD acceleration of the unconfined wind and the magnetization inferred from PWN models is known as the $\sigma$-Problem.

It was noted by \cite{coroniti1990} that allowing for finite resistivity in the wind zone, dissipation of the striped component might present a way out of this problem.  However, even when the wind-magnetisation is optimistically reduced to its non-oscillating mean value, in order to arrive at the requested $\sigma\sim10^{-3}-10^{-2}$ the pulsar needs to be very nearly perpendicular \citep[see e.g.][]{komissarov2013}.  
A more in-depth analysis of the dissipating striped wind was presented by \cite{Lyubarsky:2001} and \cite{KirkSkjaeraasen2003} who demonstrated that unless the wind multiplicity exceeds $\kappa=10^5$, due to efficient thermal acceleration out of the dissipated magnetic energy, the growth time exceeds the crossing time from the light-cylinder to the TS in the frame of the pulsar.  This somewhat high value of $\kappa$ casts some doubt that dissipation can be efficient in the wind zone.  
As a ``last exit'', \cite{2003MNRAS.345..153L} proposed that dissipation of stripes will occur directly at the termination shock, with the shock compression leading to ``driven magnetic reconnection''.  Interestingly, the jump conditions of this modified shock are identical to the case where stripes have been dissipated entirely in the upstream region.  
The PIC simulations by \cite{Sironi:2011}, and two-fluid MHD models \citep{AmanoKirk2013} in fact now show that the driven reconnection is a viable mechanism, but is able to reduce the magnetisation only down to the mean (non-oscillating) value.  

In light of this, assuming the fraction of magnetic energy in alternating stripes $\chi_\alpha(\theta)$ is dissipated entirely, straightforward calculation for the effective magnetisation yields 
\begin{align}
  \sigma(\theta) = \frac{\sigma_0\chi_\alpha(\theta)}{1+\sigma_0(1-\chi_\alpha(\theta))} \label{eq:chi}
\end{align}
where the function $\chi_\alpha(\theta)$ is given by the obliqueness angle of the pulsar $\alpha$:  From the equator where polarities cancel exactly [$\chi_\alpha(\pi/2)=0$], the striped region extends by an angle $\pm \alpha$ after which only a single magnetic polarity is present and the polar regions may remain strongly magnetised.  Assuming the wind is well described by the split-monopole solution far from the pulsar, further geometric considerations can yield the exact form of the $\chi_\alpha(\theta)$ function \citep[][]{komissarov2013}.  
Equation (\ref{eq:chi}) means that for low wind $\sigma_0$, the effective magnetisation increases linearly $\sigma(\theta)\propto \sigma_0\chi_\alpha(\theta)$ as naively expected, yet for very high wind magnetisation $\sigma_0\to\infty$ the effective value is governed by the geometry alone $\sigma(\theta)\to\chi_\alpha(\theta)/(1-\chi_\alpha(\theta))$.  While the stripe dissipation might not be sufficient to solve the $\sigma$-problem entirely, it can significantly reduce the effective flow magnetisation.   
For example, for an intermediate obliqueness angle $\alpha=45^\circ$, found by \cite{harding2008} in their modelling of the high-energy emission of the Crab pulsar, the ratio of magnetic luminosity to kinetic luminosity that is injected through the shock saturates at $L_{\rm m}/L_{\rm k}\approx0.3$ \citep[e.g.][]{PorthKomissarov2014a}.  In this simplified one-dimensional view, the pulsar obliqueness angle $\alpha$ assumes a much more prominent role than the $\sigma$-parameter of the wind.

\subsection{Termination shock}

Like the solar wind, the pulsar wind terminates when its dynamic pressure equals the thermal pressure of the surrounding shocked medium, that is the pulsar wind nebula.  
As the observations suggest very low magnetisation in the PWN, we will adopt a hydrodynamical model for some basic estimates.  A typical scale for ram-pressure balance is readily found: $r_0=\left({L}/{(4\pi p c)}\right)^{1/2}$ with $p$ the pressure of the nebular plasma.  If energy is supplied to the PWN at a constant rate $L$ and its contact discontinuity evolves according to $r_{\rm n}(t)\propto t^{\alpha_{\rm r}}$ (where we will adopt $\alpha_{\rm r}=6/5$ \citep{Chevalier1977}), adiabatic expansion of the ultrarelativistic shocked PWN implies $\dot{E} = L-\alpha_{\rm r}{E}/{t}$.  Thus the energy accumulated over time in the PWN bubble is
\begin{align}
E= \frac{L t}{1+\alpha_{\rm r}}
\end{align}
with the initial condition $E(0)=0$. Assuming a uniform distribution of the PWN pressure $p=E/(4\pi r_{\rm n}^3)$ we find the scale of the TS
\begin{align}
r_0 = \frac{(1+\alpha_{\rm r})}{\alpha_{\rm r}} r_{\rm n}(v_{\rm n}/c)^{1/2}
\label{eq:rtshydro}
\end{align}
Where $v_{\rm n}$ is the observed expansion speed of the nebula. In case of Crab, $v_{\rm n}\approx 2000 \rm km\, s^{-1}$ and using the observed nebula radius of  $r_{\rm n}=1.65\rm pc$ \citep{hester2008} we find the values $r_0=0.1\, r_{\rm n}=0.17\rm pc$.  
The very good match of this estimate with the extent of the X-ray inner ring $r_{\rm ir}\simeq 0.14\rm pc$ \citep{weisskopf2000}, seems too good to be coincidental which is why the inner ring is often identified directly with the location of the TS.   
In terms of the light cylinder we have $r_0=5\times 10^8 \varpi_{\rm LC}$ which means that both rotation and radial magnetic field  are sub-dominant at the scale of the TS.  We will hence proceed in the ``toroidal approximation'': $v_\phi/c\approx 0$, $B_r/B_\phi \approx0$.  

Given the anisotropy of the wind power (\ref{eq:michelpower}) and (\ref{eq:obliquesr}), the TS will attain pressure equilibrium at different locations:  
near the poles, the wind power steeply declines and the shock retreats back towards the pulsar.  In the equatorial plane where the wind power is maximal on the other hand, the shock will bulge out further.  This mechanism provides a simple explanation to the torus structure observed in many PWN \citep{lyubarsky2002}.   In Section \ref{sec:knot}, more detailed calculations concerning the shape of the oblique shock and radiative signatures of particles emitting close to the shock will be discussed.  

Depending on their local wind magnetization, streamlines that pass over the shock behave qualitatively very different.  
In high-$\sigma$ regions i.e. close to the axis, the flow can remain highly relativistic with downstream Lorentz factors larger than $\sigma^{1/2}$, the value obtained for the perpendicular shock.  For $\sigma\to\infty$, streamlines simply continue to flow out radially after traversing the shock.  In finite-$\sigma$ regions with non-vanishing shock compression, the force-free equilibrium of the wind is destroyed and flow-lines start curving towards the pole due to the pinch-force of the magnetic field.  This mechanism of jet formation due to the magnetic hoop-stress has been observed in axisymmetric \citep[e.g.][]{ssk-lyub-03} and full 3D simulations \citep[e.g.][]{PorthKomissarov2014a}.  
In the low-sigmam, e.g. striped regions, due to the conservation (compression) of the tangential (normal) velocity fields, streamlines experience shock aberration and instead get  deflected towards the equatorial plane.  This leads to the occurrence of ``wisps'' emitted from the shock to be further discussed in Section \ref{sec:wisps}.

\subsection{Nebula}
\label{sec:nebula}
The stationary solutions of the system (\ref{eq:continuity} - \ref{eq:induction}) in the toroidal approximation were exhaustively studied by \cite{kc84a}.  They showed that assuming the wind is ultra relativistic $\Gamma_1\gg1$ and expands adiabatically after encountering a strong shock, e.g. $\Gamma_1\gg\Gamma_2$, $\hat{\gamma}=4/3$, the flow can be parametrised entirely in terms of the upstream $\sigma$ (here subscripts 1 and 2 indicate the upstream and downstream medium, respectively). 
The asymptotic flow velocity in the shocked nebula then becomes
\begin{align}
v_{\infty} = \frac{\sigma}{1+\sigma} c
\end{align}
and hence it is clear that the boundary condition $v_{\rm n}\approx 2000 \rm km\, s^{-1}$ could only be satisfied by requiring low values of $\sigma<10^{-2}$.  
Going further, assuming we indeed have a weakly magnetised wind, the large sound-speed in the shocked bubble will quickly equilibrate density and pressure and thus (\ref{eq:continuity}) implies $v_r\propto r^{-2}$ after crossing over the termination shock (TS).  Then, from (\ref{eq:induction}), we find $B_\phi\propto r$.  The initially sub-dominant magnetic energy increases rapidly until equipartition with the thermal energy is obtained.  From then on, since $v_r(r)>v_{\infty}$ the velocity approaches a constant value and the magnetic energy decreases again $B_\phi\propto 1/r$.  
This dynamic behaviour further tightens the limit on $\sigma$ in the toroidal approximation, e.g. \cite{kc84a} obtained a best fit to the observed boundary conditions of Crab nebula for $\sigma\simeq 0.003$.  

An additional argument for low magnetisation of the injected wind was put forward by \cite{rees-gunn-74}:  ignoring the spin-down of the pulsar for the moment, each turn adds one ``loop'' of magnetic field and magnetic flux conservation implies $B_\phi\sim t$.  Thus for the magnetic energy accumulated in the nebula we have $\mathcal{E}_{\rm mag}\sim t^2$.  At the same time, particle energy is injected at constant rate $\mathcal{E}_{\rm part}\sim t$. In order to arrive at todays approximate equipartition $\mathcal{E}_{\rm mag}\sim\mathcal{E}_{\rm part}$ (and accounting for nebula expansion and spin down of the pulsar), \cite{rees-gunn-74} estimate a wind magnetisation of $\sigma\approx0.01$.

The elegance and simplicity of the \cite{kc84a} model (KC) and the \cite{rees-gunn-74} argument renders KC models a standard tool to recover parameters of observed PWN \citep[e.g.][]{sefakodeJager2003,PetreHwang2007} although there are significant problems in reproducing the X-ray spectral index maps of several PWN \citep{reynolds2003,tang2012,NynkaHailey2014,PorthVorster2016}.  
These deficiencies, the uncomfortably small value of $\sigma$ and the obviously non-spherical \emph{jet and torus} morphology revealed by the Chandra X-ray telescope \citep{KargaltsevPavlov2008} make a strong case for multi-dimensional models of PWN.  
To date, a number of groups have carried out axisymmetric relativistic MHD simulations of pulsar wind nebulae \citep{ssk-lyub-03,ssk-lyub-04,del-zanna2004,bogovalov2005,camus2009,Olmi:2014}. Although rather different computer codes were employed, all these simulations produce quite similar results: the numerical solutions reproduce well the observed jet-torus structure and suggest dynamical explanations to the wisps (Section \ref{sec:wisps}) and curious inner knot (Section \ref{sec:knot}).

Despite these successes, the axisymmetric models can not significantly lift the tension on the wind magnetisation parameter:  if $\sigma>0.1$, the termination shock is pushed far back to the pulsar and excessively strong jets develop that even punch through the nebula bubble \citep[e.g.][]{PorthKomissarov2013}.  
Both effects are linked to the accumulating hoop stress caused by conservation of the toroidal magnetic flux.  
In order to accommodate wind $\sigma\approx1$ with the observations, two additional ingredients are essential:  1. Destruction of hoop-stress via fluid instabilities (no more toroidal approximation) and 2. magnetic dissipation in the nebula volume.  

The fact that the toroidal (z-pinch) equilibria used in the axisymmetric modelling \citep[e.g.][]{begelman1992} are unstable to the MHD kink instability was first discussed by \cite{begelman1998}.  
The authors speculated that the disrupted configuration may be less demanding on the magnetization of pulsar winds. Indeed, one would expect the magnetic pressure due to randomized magnetic field to dominate the mean Maxwell stress tensor, and the adiabatic compression to have the same effect on the magnetic pressure as on the thermodynamic pressure of relativistic gas. Under such conditions, the global dynamics of PWN produced by high-$\sigma$ winds may not be that much different from those of PWN produced by particle-dominated winds. These expectations have received strong support from numerical studies \citep{Mizuno:2011aa} of the magnetic kink instability for the cylindrical magnetostatic configuration.  These simulations have shown a relaxation towards a quasi-uniform total pressure distribution inside the computational domain on a dynamical time-scale. 

In addition to the dissipation of magnetic stripes in the wind or at the termination shock, the magnetic dissipation could occur inside the PWN as well \citep{lyutikov2010c,komissarov2013}. In fact, the development of the kink instability near the axis and Kelvin-Helmholtz instabilities operating in the equatorial region \citep{camus2009} are bound to facilitate such dissipation. In principle, simultaneous observations of both the synchrotron and inverse-Compton emission allow a measurement of the energy distribution between the magnetic fields and the emitting electrons (and positrons). From a simple ``one-zone'' model of the Crab nebula, it follows that its magnetic energy is only a small fraction, $\sim1/30$, of the energy stored in the emitting particles (\cite{MeyerHorns2010}; \cite{komissarov2013}). Thus unless the striped zone fills almost the entire wind volume, additional magnetic dissipation inside the nebula is required.

\subsection{Insights from 3D simulations of PWN}

With the discussion of the previous sections, the setup of a PWN simulation becomes straight-forward:  
first the total energy flux of the wind $f_{\rm tot}(r,\theta)$ needs to be chosen. Typically the energy flux is assumed to follow the Poynting flux of the aligned (\ref{eq:michelpower}) or perpendicular rotator (\ref{eq:obliquesr}), however more sophisticated formulas for oblique rotators are now also being applied \citep[see][]{BuhlerGiomi2016}.  
Then the striped zone is modelled by choosing an appropriate function for $\chi_\alpha(\theta)$.  Here the prescriptions of various groups differ somewhat (c.f. Eq. (5) of \cite{PorthKomissarov2013}, and Eq. (2) of \cite{Olmi:2014}).  With the effective wind-magnetisation given by Equation (\ref{eq:chi}), the magnetic and kinetic energy fluxes follow to
\begin{align}
f_{\rm m}(r,\theta) = \sigma(\theta)\frac{f_{\rm tot}(r,\theta)}{1+\sigma(\theta)};\ \ f_{\rm k}(r,\theta) = \frac{f_{\rm tot}(r,\theta)}{1+\sigma(\theta)}\, .
\end{align}
Adopting the wind Lorentz-factor $\Gamma$, one can then solve for the density and magnetic field strength of the cold wind ($p\ll\rho$) 
\begin{align}
\rho(r,\theta) = f_{\rm k}(r,\theta)/(\Gamma^2c^2v_{r});\ \ B_{\phi}(r,\theta)=\pm\sqrt{4\pi f_{\rm m}(r,\theta)/v_r}
\end{align}
where we have taken advantage of the toroidal approximation.  Note that the magnetic field still reverses polarity from the northern to the southern hemisphere which gives rise to a magnetic null line within the nebula bubble.  Due to numerical reasons, a realistic Lorentz-factor of $\sim10^5$ cannot be realised in the simulations which typically set $\Gamma\sim10$.  In the hydrodynamic regime, it can be shown (e.g. \cite{ssk-lyub-04}) that the jump conditions of the ultrarelativistic shock are fairly well approximated already at this low value of $\Gamma$.  
Finally, the outer nebula boundary is modelled by assuming a supersonically expanding hydrodynamic shell of stellar ejecta with velocities at the inner edge $\sim1000\rm km s^{-1}$ (e.g. \cite{del-zanna2004}).  

\begin{figure}[htbp]
\begin{center}
\includegraphics[width=0.9\columnwidth]{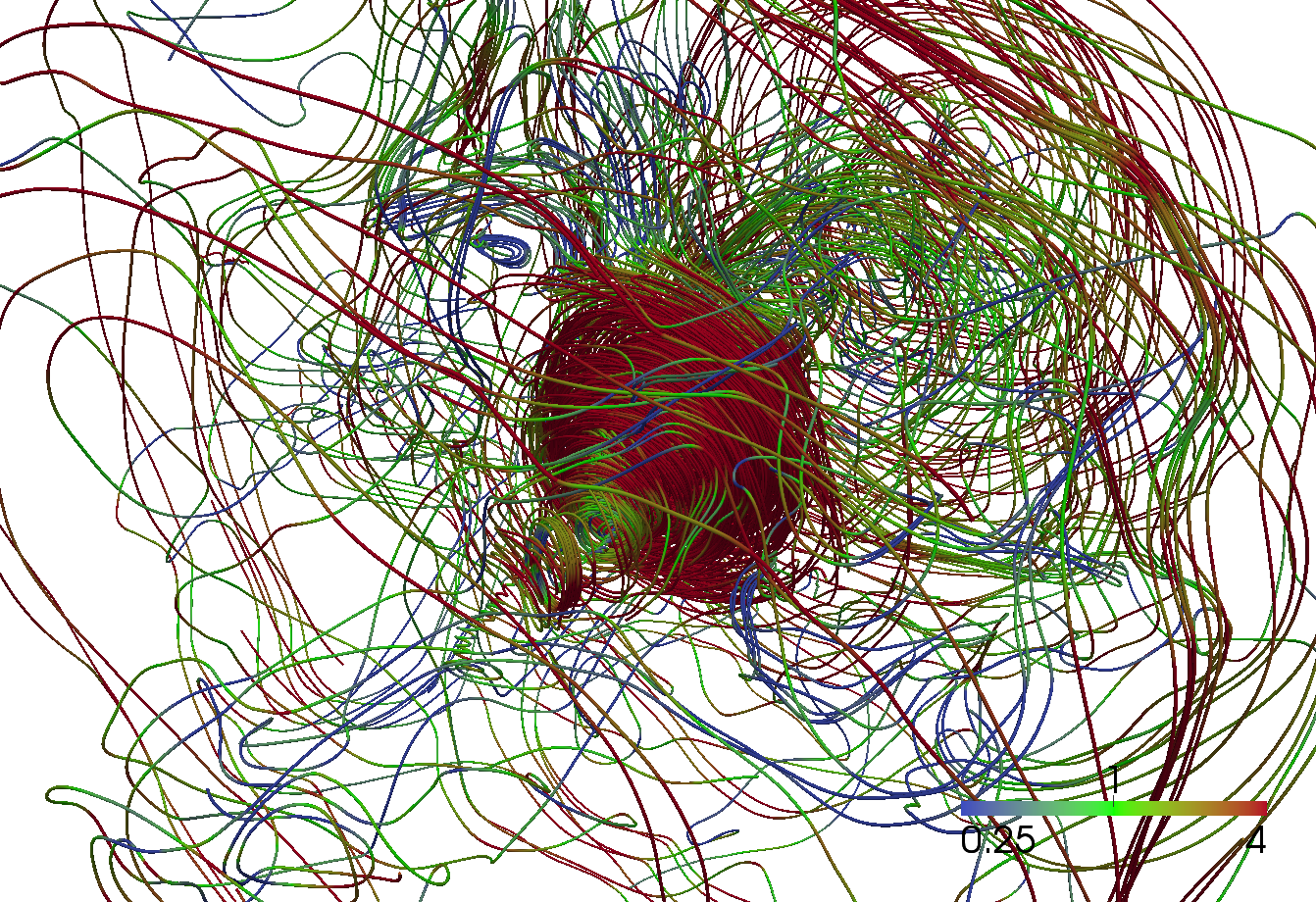}
\caption{{\label{fig:spaghetti} Rendering of magnetic field lines in 3D simulations of PWN.  Color indicates $B_\phi/B_{p}$: in the innermost regions, toroidal field (red) is injected which randomises further out and develops a poloidal component (blue).  The kink instability of the leads to the corkscrew morphology of the inner jet.
}}
\end{center}
\end{figure}

Contrary to axisymmetric simulations, the 3D case allows the magnetic field to deform and develop a poloidal component.  Since injection of fresh toroidal field competes with fluid instabilities in the downstream flow, whether the telltale jet-and-torus structure can also be established in 3D is not entirely obvious.  
In Figure \ref{fig:spaghetti}, a rendering of field lines from 3D simulations illustrates the process.  Indeed, toroidally dominated field is present only in the direct vicinity of the TS.  
The strong jets observed for $\sigma=1$ in axisymmetric simulations do not survive in 3D. In fact, the jet rapidly becomes unstable to the $m=1$ kink instability as predicted by \cite{begelman1998}.  Nonetheless, outflow velocities in the polar plume reach $\approx0.7c$, similar to the pattern speeds observed in Crab ($\sim 0.4c$) \citep{hester2002} and Vela (0.3c-0.7c) \citep{PavlovTeter2003}.

\begin{figure}[htbp]
\begin{center}
\includegraphics[width=0.98\columnwidth]{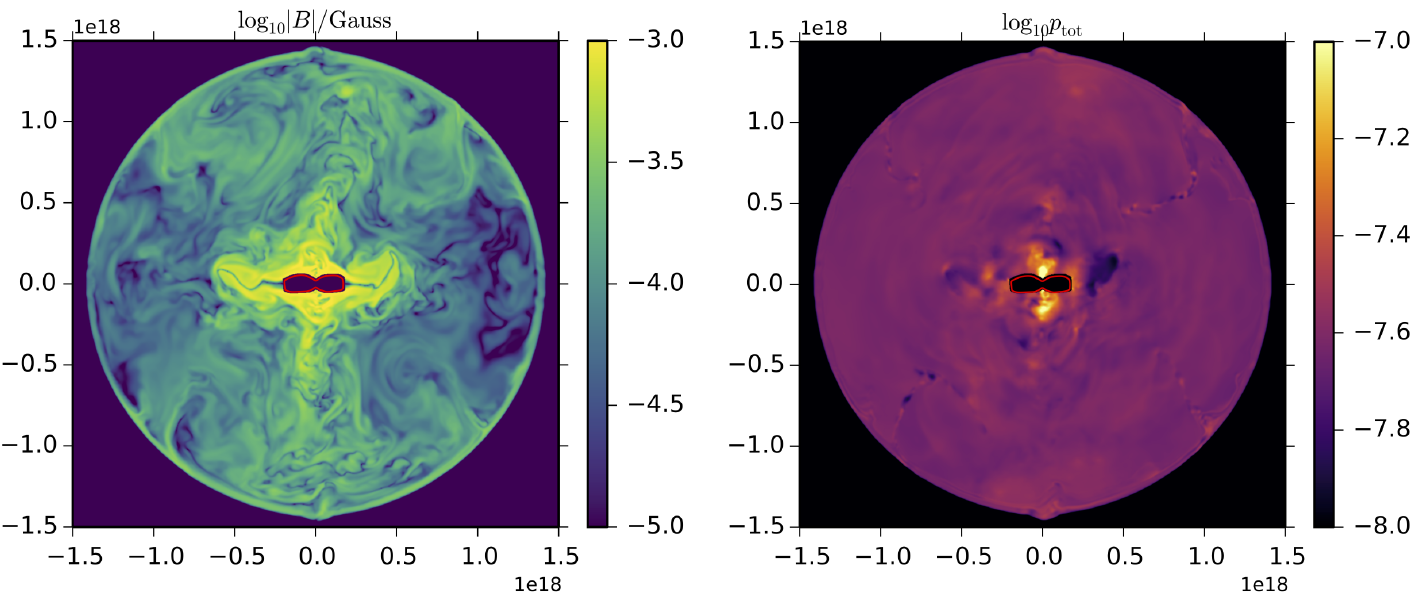}
\caption{{\label{fig:slices} Global views of the nebula showing magnetic field strength (left) and total pressure in the nebula (right).  There is a significant range in magnetic field strength, with $|B|\approx1\rm mGauss$ near the TS and a volume-average of $\approx 100\, \mu\rm Gauss$.  Apart from the hotspot in the dissipation region of the polar beam, the total pressure is close to uniform in the nebula.
}}
\end{center}
\end{figure}

An overview of the large-scale dynamics in the 3D models is given in Figure \ref{fig:slices}.  As the z-pinch is rapidly destroyed, the total pressure is now more or less uniform and resembles the purely hydrodynamic models even for highly magnetised wind.  
Only at the base of the plumes flow compression is significantly enhanced.  
Because of the dramatically altered magnetic field distribution in the nebula, the termination shock does not dive back towards the pulsar and values compatible with the hydrodynamic prediction leading to Eq. (\ref{eq:rtshydro}) could be found even for $\sigma_0=3,\alpha=45^\circ$ \citep{PorthKomissarov2014a}.  
The magnetic field strength exhibits a large range with maximal values in the ``arch flow '' just on top/below the oblique TS.  As long as axisymmetry is approximately conserved, the magnetic field strength follows the scalings discussed in Section \ref{sec:nebula}, hence the magnetic energy density increases until reaching equipartition with the plasma pressure.  Throughout the torus region, the magnetic field remains strong and toroidally dominated, however fluid instabilities in the fast equatorial shear flow lead to turbulence starting at $\approx 2 $ shock radii.  It is interesting to note that the shock compresses streamlines that carry magnetic field with opposing polarity towards the equator, turning the null-line into a large-scale current sheet.  It is in this current sheet where the majority of magnetic dissipation takes place, a fact that was already observed in 2D simulations of \cite{camus2009}.

\begin{figure}[h!]
\begin{center}
\includegraphics[width=0.42\columnwidth]{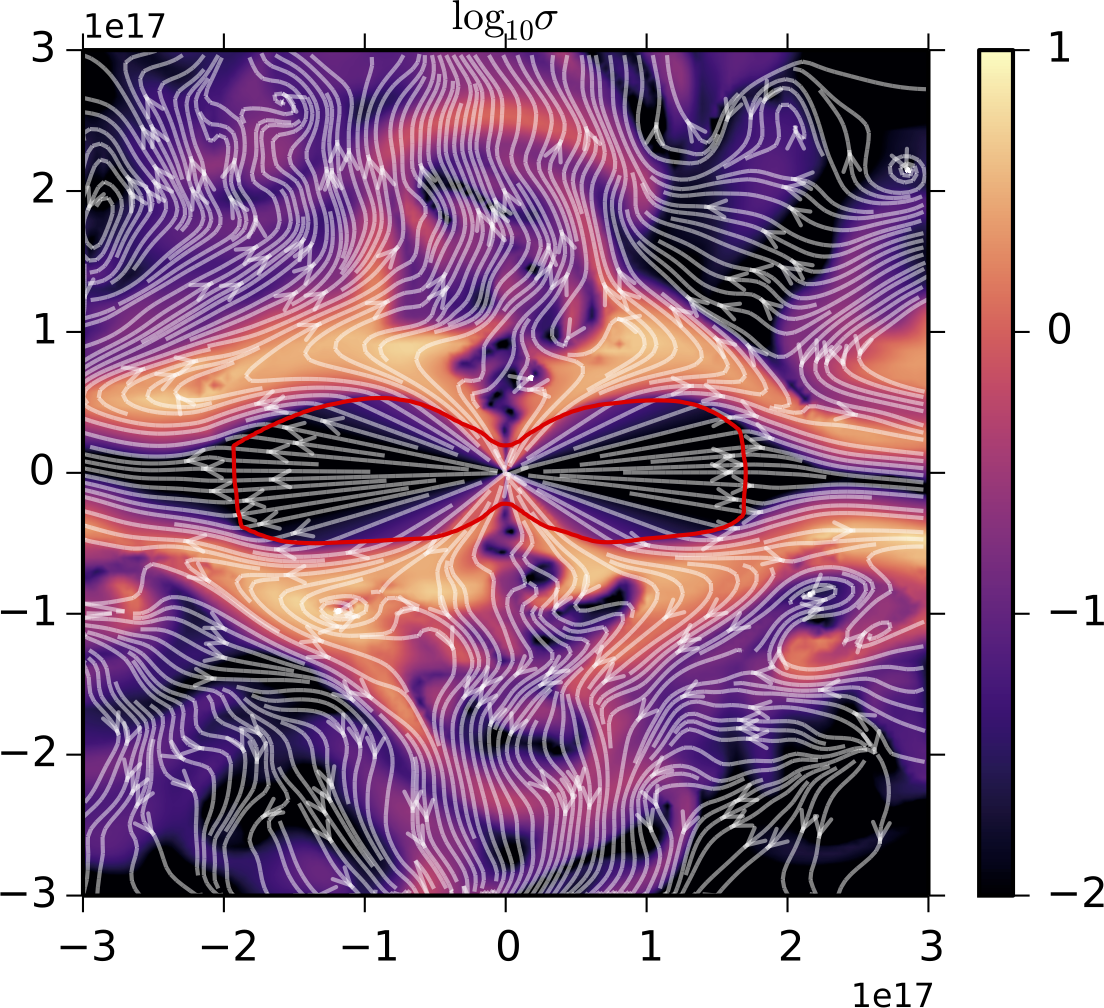}
\caption{{\label{fig:rmhdflow} Zoomed in view on the termination shock (red contour) showing the logarithm of the magnetization.  Parameters of the simulation are:  $\sigma_0=1$, $\alpha=45^\circ$. In-plane streamlines (white contours) get deflected at the shock and turn towards the equator or polar axis, depending on the magnetization.  Note that the magnetization in the nebula can be significantly higher by a factor $\sim10$ than the one injected with the wind.
}}
\end{center}
\end{figure}

A  closeup view of the TS is shown in Figure \ref{fig:rmhdflow} illustrating a typical flow field.  In the vicinity of the TS, the predictions from axisymmetric calculations remain largely valid:  we obtain a separation of flow lines into equatorial and polar flow.  Velocities in the fast equatorial shear flow reach up to $\approx0.5c$.  Re-focussing of flow lines from the polar regions forms a plume-like vertical outflow and an inner highly unstable ``polar beam''.  Due to the kink instability of the polar beam, the actual formation of the jet is offset from the TS -- in good agreement with the observations.  This dynamical behaviour also suggests an identification of the highly variable ``Sprite'' (see e.g. \cite{hester2002}) located at the base of the jet with the violently unstable polar flow.

It is interesting to note that (similar to the 1D KC84 model) due to expansion and deceleration of streamlines carrying a toroidal field, the magnetisation of the flow in mid-latitude regions can in fact increase over the value injected at the TS.  The high $\sigma$ regions in the nebula volume might present suitable locations for rapid particle acceleration in connection with the Crab flares \citep[e.g.][]{LyutikovSironi2016}.  Recent advances in modelling the elusive Crab flares will be discussed further in Section \ref{sec:flares}.

\subsection{Radiative predictions from particle transport models}
\label{sec:radiation}

Radiative predictions are necessary to ultimately test our theoretical models and simulations with respect to observed systems. While the general non-thermal emission processes in PWN were rapidly identified, namely synchrotron radiation from non-thermal electrons and inverse Compton scattering of ambient radiation by the very same electrons \citep{shklovsky:1953,dombrovsky1954nature,atoyan1996}, ever increasing spatial and temporal resolution ask for more and more refined models of the PWN emission.
In their core, current emission models follow the suggestion of \cite{kc84b}, that non-thermal electrons are accelerated at the TS and then follow streamlines in the nebula where they experience adiabatic and radiative losses.  In the optical and X-ray bands where the cooling timescale is comparable to the crossing time, this promotes a ``center-filled'' appearance and gradual steepening of the X-ray spectral index, observed especially in young PWN \citep[e.g.][]{SlaneChen2000,SlaneHelfand2004}.  
Sophisticated methods to track the distribution of non-thermal particles injected at the TS via passive tracer scalars were devised \citep{delzanna-06,camus2009} and are now standard practice in MHD simulations of PWN.
While acceleration at the TS appears as a good working assumption, the details of the particle acceleration process are far from understood and the large inferred acceleration efficiency appears in conflict with the superluminal nature of the shock in the framework of standard models.  
This will be discussed further in Section \ref{sec:wispacc} where the observational signatures of various acceleration sites on the shock are compared.

\begin{figure}[h!]
\begin{center}
\includegraphics[width=0.98\columnwidth]{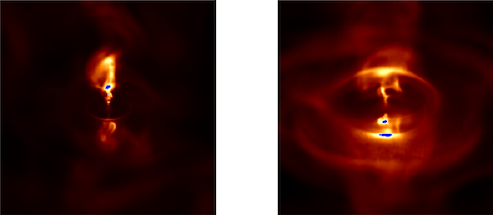}
\caption{{\label{fig:maps} Synthetic emission maps assuming energy of the non-thermal particles scales as  $\epsilon\propto p$,  i.e. assuming only adiabatic effects.  Two simulations are compared: $\sigma_0=3,\alpha=10^\circ$ (left) and $\sigma_0=1,\alpha=45^\circ$ (right).  In high-magnetisation, low obliqueness models, the torus and wisp regions are under-luminous and the emission from the jet (in fact the polar beam) dominates.  Morphology and parameters of the right-hand model fit well to the optical images of the Crab PWN.  Blue color indicates where pixels were saturated ($80\%$ peak intensity) in order to better show faint features.
}}
\end{center}
\end{figure}

On the qualitative level, synthetic maps of synchrotron emission from PWN are able to reproduce a stunning amount of features: foremost the famous jet-and-torus morphology whose dynamical origins were previously discussed, but also finer details seen primarily in the high resolution observations of the Crab nebula.  We highlight the aforementioned sprite, the inner knot (sections \ref{sec:knot}) and the wisps (Section \ref{sec:wisps}).
As the details of energy injection into the PWN are intimately related to fundamental parameters of the rotating neutron star, one can attempt to constrain pulsar parameters with the morphology of the synchrotron emission in the nebula \citep[e.g.][]{BuhlerGiomi2016}.  
In this vein, in Figure \ref{fig:maps}, we show synthetic emission maps of the region around the TS from two 3D simulations: one with parameters $\sigma_0=3,\alpha=10^\circ$ (left) and one with $\sigma_0=1,\alpha=45^\circ$ (right).  While the right-hand panel provides a good match to the morphology of Crab, the torus structure is entirely suppressed in the left-hand panel which might find its likeness in one of the jet-dominated sources \citep[][]{KargaltsevPavlov2008}.  This result is not surprising: as the extent of the striped zone is decreased, more streamlines are re-focussed into the jet (cf. figure 14 of \cite{PorthKomissarov2014a}).  The recent 3D simulations presented by \cite{Olmi2016} confirm this finding.  Hence the jet/torus flux ratio could yield a valuable handle on the pulsar obliqueness.  

On the quantitative level, some progress could be made recently with detailed modeling of the inner knot of Crab nebula by \cite{YB-15} and \cite{LyutikovKomissarov2016}.  If the knot-feature is caused by beamed emission right behind the TS, then the local magnetisation must be $\sigma<1$, otherwise the observed properties of the knot cannot be reproduced.  For an inclination of the Crab pulsar of $60^\circ$ with respect to the plane of the sky and an assumed obliqueness angle of $45^\circ$, the knot falls into the striped zone and low post-shock magnetisation is in fact expected (see Section \ref{sec:knot} for details).  

Moving away from the TS, the uncertainties in radiative models increase:  currently, the synthetic maps tend to over-predict the contrast between torus and ambient emission, as well as the optical and radio polarization degree.  For example, the unresolved polarization degree of the Crab nebula in the optical band is $\Pi=9.3\pm0.3\%$ \citep{OortWalraven1956}, whereas the simulations produce an over three times higher value.
Both effects are related, e.g. with decreasing emissivity in the ordered-field torus region as compared to the turbulent bulk, the average polarization will also decrease.  
This points at several potential shortcomings of the current 3D models:  
1. Their short duration does not allow the dense filaments due to the Rayleigh-Taylor instability to develop and penetrate deep into the nebula (e.g. \cite{porthRT2014}), an effect that would likely increase the randomization of magnetic field.  
2. The finite resolution global simulations might over-predict the dissipation of magnetic field in the bulk.
3. The prescription of  localized acceleration, occurring at the TS only, as in the KC84 original model, is possibly overly simplified and particle re-acceleration in the bulk might have to be taken into account.
Future higher resolution and longer duration 3D simulations will need to address these questions.  

Due to the inherent complexity of the data, emission maps from dynamical simulations are not well suited to yield quantitative information of PWN, and steps to reduce the dimensionality must be taken so that parameters can be obtained via model fits.  In this sense, \cite{volpi2008} fitted the spectral energy distribution (SED) of axisymmetric MHD simulations to the Crab nebula, taking into account two species of synchrotron emitting particles (radio- and optical/Xray-electrons) and realistic target photon fields for the IC emission.  
The result can be understood as a different flavour of the $\sigma$-problem:  
In order to reproduce the high-energy jet/torus morphology and due to the strong axial compression, the average field turns out to be $3-4$ times smaller than predictions from KC84 models \citep{de-JagerHarding1992}.  Hence in order to arrive at the observed synchrotron flux, the electron density must be increased accordingly. This is problematic not only as it adds to the tension on the multiplicity parameter $\kappa$, but also since it leads to an exaggerated IC emission \cite[see also the discussion in][]{Amato:2014}.  
With their more uniform distribution of the magnetic field even for high magnetisation, 3D models might present a way out of this dilemma and research in this direction is ongoing \citep{Olmi2016}.

An alternative technique to the advection of passive tracer scalars was followed by \cite{PorthVorster2016}: here, X-ray synchrotron electrons are treated as test-particles that experience the Lorentz force due to the MHD background fields.  This allows to directly map out the advection and diffusion of particles embedded in the PWN flow.  It was found that due to large velocity fluctuations, the transport exhibits a diffusive character and a typical diffusion coefficient of 
\begin{align}
D_{\rm Ls} \sim \frac{1}{3} v_f L_{\rm s} = 2.1 \times 10^{27}\left(\frac{v_f}{0.5c}\right)\left(\frac{L_{\rm s}}{0.42\rm Ly}\right)\ \rm cm^2~s^{-1}\, \label{eq:DE}
\end{align}
is suggested following the turbulence with driving scale $L_{\rm s}$ -- the size of the termination shock, and a typical velocity at this scale $v_f$.  
Note that the value of $D_{\rm Ls}$ for Crab nebula becomes very similar to the one estimated by \cite{AmatoEtAl00} and also to that obtained in the early model due to \cite{WeinbergSilk1976} of $1.9 \times 10^{27}\rm cm^2 s^{-1}$.  
As particles are mostly following the velocity field, e.g. the  first order drift velocity in the toroidally dominated regions $\mathbf{v}_D=\mathbf{E\times B}/B^2$ just corresponds to the flow velocity in the poloidal plane $\mathbf{v}_D=\mathbf{v}_p$, the diffusive transport is independent of the particle energy, as assumed already in the models of \cite{gratton1972,Wilson1972}.  Further discussion on diffusion in PWN can be found in \citep{tang2012}. 
 With the spherically averaged MHD simulation as background, the particle transport taking into account advection, diffusion and synchrotron losses, can be fitted to X-ray observations of three young PWN confirming that diffusion makes an important contribution in PWN. In particular, the model of \cite{PorthVorster2016} yields better fits to the gradual increase of the X-ray spectral index than a traditional KC84 model.

\section{Wisps as probes of the particle acceleration mechanism in the Crab nebula}
\label{sec:wisps}
PWN are among the most powerful accelerators in the Galaxy, with the Crab showing an acceleration efficiency of order 20\% and producing particles up to PeV energies. 
It is a common opinion that the acceleration of those high energy particles might take place in the proximity of the wind termination shock. However the termination shock is a very hostile environment for acceleration, due to its nature of magnetized and relativistic shock. 
Relativistic shocks have indeed been proven to be efficient accelerators only in two cases: when the magnetic field is quasi parallel to the shock normal, in particular when it is sub-luminal (meaning that the angle between $\vec{B}$ and $\vec{v}_\mathrm{shock}$ is $\theta\ll\theta_c\simeq 1/\gamma_\mathrm{shock}$), or when the shock is poorly magnetized, with the magnetization $\sigma$, defined as the ratio of Poynting flux to particle kinetic energy flux in the wind, below $\sim 10^{-3}$.

The observed spectrum of the Crab nebula, similar to other PWNe, seems to suggest different acceleration mechanisms for low and high energy emitting particles, since it is a broken power law $N(E)\propto E^{-\gamma_e}$, with $\gamma_e\sim 1.5$ at low energies (below $\sim 100$ GeV) and $\gamma_e\sim 2.2$ at higher ones (up to a few TeV).

At present, two main candidates have been invoked as possible mechanisms for particle acceleration at the Crab TS.

The slope of the radio component is compatible with driven magnetic reconnection. 
This mechanism was largely investigated by \citep{Sironi:2011} and \citep{Sironi:2013}, who have performed numerical simulations compatible with the Crab nebula's shock conditions. 

They assume that the striped morphology of the pulsar wind is maintained all the way to the TS, where the compression causes the magnetic field to reconnect. 
Reconnection islands form in the flow and particles can be accelerated by the un-screened electric field. 
The properties of the resulting spectrum depend on the flow magnetization and on the ratio between the wavelengths of the stripes and the particle Larmor radii. 
This ratio can be connected to the value of the pulsar pair multiplicity $\kappa$, and the result in the case of the Crab nebula is that, in order to reproduce the observed radio spectrum, a magnetization of $\sigma\gtrsim 30$ and a multiplicity of $\kappa \gtrsim 10^8$ are needed. We remind the reader that the already mentioned pulsar multiplicity $\kappa$ is related to the efficiency with which particles are generated in the pulsar magnetosphere, namely it is defined as the number of pairs produced by a single primary electron that emerges from a polar cap of the pulsar. 

The main problem with this scenario is that it is very difficult to explain such a high value of the multiplicity: from observations and theoretical models we expect much lower values, with $\kappa\sim few \times 10^5$ at most \cite{timokhin15}. 
The only possibility to reduce the required value of $\kappa$ is to accelerate particles near the polar cusps of the TS, where the shock front is nearer to the pulsar, and as a consequence the particle density is much higher (given that in the wind it decreases as $\approx 1/r^2$). 
On the other hand, no stripes are expected to be present at these locations, and reconnection can only be obtained due to O-point dissipation.

Moreover, a wind with the required high number of pairs, $\kappa\sim10^8$, is expected to reconnect well before the TS \citep{Lyubarsky:2001}, causing the magnetization to lower well below the required minimum for the mechanism to be viable \citep{Amato:2014}.

A different mechanism can be invoked to account for the much steeper high energy component of the spectrum. A photon index of $\gamma_e \simeq 2.2$ is in fact what one expects from Diffusive Shock Acceleration (DSA), i.e. Fermi I like acceleration at relativistic shocks. 
Contrary to driven magnetic reconnection, DSA is effective only if the magnetization is sufficiently low, in particular $\sigma \lesssim 10^{-3}$ is required \citep{Sironi:2009}. 
MHD simulations taught us that, if the emission from Crab has to be reproduced, this condition can only be realized in a thin latitude strip around the pulsar equator or in the vicinities of the polar axis, while at intermediate latitudes the flow magnetization must be substantially larger than this, likely $\sigma$ of order a few \citep{komissarov2013}. 
Therefore, at least DSA can only be effective in a small sector of the shock, where magnetic dissipation is sufficiently strong so as to ensure a local condition of quasi-unmagnetized plasma \citep{Amato:2014}.

The other possibility that has received some attention in the literature is resonant absorption of ion-cyclotron waves in a ion-doped plasma \citep{HoshinoEtAl92,AmatoArons06}. This requires that most of the energy of the wind is carried by ions and its viability does not depend on the value of the magnetization. 
The idea is that pairs were accelerated by resonant absorption of the cyclotron radiation emitted by ions in the wind, which are set into gyration by the enhanced magnetic field at the shock crossing \citep{Amato:2014}.

At present there are no indications for significant role of ions in the pulsar wind energetics, or that ions are present at all.  The observations of high energy neutrinos provides the primary tool to assess this matter (e.g. \citet{2003A&A...402..827A}). While current limits, set by 6 years of observations with IceCube, are still not very stringent, upcoming neutrino facilities are promising to either detect neutrinos from a few bright PWNE or to constrain the fraction of pulsar wind energy flux carried by ions to levels below 10\% \citep{2016arXiv160501205D}.  
Given the current uncertainties, however, the possibility that electron acceleration occurs through resonant absorption of ion cyclotron waves will not be discussed further in this work.

\subsection{Wisps from Crab}

The inner region of the Crab nebula has been known to be highly variable at optical wavelengths since the late 60s, when \cite{Scargle:1969} first identified some bright, arc-shaped and variable features, that he named ``wisps''. 
They appear to be periodically produced at about the expected location of the TS, which can be seen as the first bright ring surrounding the central dark region of the pulsar wind, and then move outwards with mildly relativistic velocities, as one expects for a post-shock hydrodynamic flow. 
Their typical period between appearance close to the TS and disappearance in the bulk of the nebula can be of the order of days, months or even years, depending on the energy band in which they are observed (X-rays, optical or radio respectively).
They also appear to be more prominent in the north-west sector than in the south-east.

Few years after the first wisps identification, the properties of such features were extensively studied at different wavelengths, in particular focusing on radio and optical bands, and more recently also on the X-ray band. 
The most important conclusion was that wisps at different wavelengths are not coincident and that they are seen to propagate at various speeds \citep{Bietenholz:2004,Schweizer:2013}.
The observed discrepancies between radio and optical wisps led \cite{Bietenholz:2004} to conclude that the two populations must have a different acceleration mechanism and/or site, and possibly the same conclusion might be drawn for X-ray wisps.

\subsection{Wisps as probes of the particle acceleration}
\label{sec:wispacc}

As we discussed previously, the different acceleration mechanisms proposed require very different physical conditions to be viable, which are not realized everywhere along the shock surface. Identifying the location at which particles are accelerated can then put strong constraints on the acceleration mechanism at work.

One way to test different scenarios of acceleration is by comparison with observations of the variability they entail in the inner nebula at various frequencies: since wisps are seen to start so close to the TS, at least at optical and X-ray frequencies, where radiative losses are important, they trace freshly injected particles, and in the simulated maps, their appearance and motion depends on the location at which the emitting particles are injected in the nebula.

In \cite{Olmi:2014} it was shown that wisps' properties at radio wavelengths are well reproduced without invoking any \emph{ad hoc} mechanisms, with the bulk flow of the nebula acting as the main driver for the observed wisps appearance and motions. 
In MHD models, wisps appearance is thus totally due to the combined effects of the locally enhanced magnetic field, just downstream of the TS, and the Doppler boosting effect, since channels with significant $v/c$ form along oblique sectors of the shock surface.
In particular, Doppler boosting is responsible for the angular profile of the wisps, as well as for the enhanced brightness of the front side of the nebula with respect to the back side, that appears to be very faint. 
The intensity contrast between distinct wisps is, on the contrary, strongly connected to the local magnetic field strength.

Within an MHD description, the fact that wisps arise at different wavelengths with different properties (different locations and outward velocities) suggests a difference in the acceleration sites of the particles responsible of such emission.
In fact, assuming that the emitting particles are all accelerated at the TS, including the radio component, the discrepancies can only be explained by choosing different acceleration regions along the TS for distinct distributions. 
If particles with different energies are injected at different latitudes along the TS, the paths induced by the post-shock flow structures, and the adiabatic and synchrotron losses, will also be different, and features at different energies are expected to be not coincident, as observed.

In \cite{Olmi:2015} axisymmetric relativistic  MHD simulations are used to constrain the acceleration sites of particles responsible for the observed wisps at the different frequencies in the Crab nebula. 

Considering several different scenarios, with particles of different energies being injected in different sectors of the TS, with polar and equatorial cones defined with various angular extents, wisp properties are extracted on top of the simulated emission maps at various wavelengths.

The entire Crab nebula is simulated up to its present age ($\sim 1000$ yr), with outputs sampled with monthly frequency during the last 10 years of evolution. 
In analogy with the observation based study carried out by \cite{Schweizer:2013}, wisp intensity profiles are extracted from radio, optical and X-ray emission maps from a $3^{\prime\prime}$ wide slice centered on the polar axis in the upper hemisphere of the nebula, where wisps are seen to be more prominent.
For ease of comparison with observational data, profiles are also convolved with the appropriate instrumental PSF (relative to the instrument used for the data considered for comparison in each band) and only intensity peaks with $I\geq I_\mathrm{max}/3$ are taken into account, where $I_\mathrm{max}$ is the maximum value of the intensity in each map. 
This cut off is applied in order to remove the background of weaker variations, that are not useful for the comparison with data. 

This procedure is then repeated for each output in the considered interval of 10 years, and for each one of the defined cases for particles injection, namely:
\begin{enumerate}
\item  particles are injected uniformly along the entire shock front, for all the three families;
\item  particles are injected in a wide equatorial sector along the shock ($\theta \in \left[ 20^\circ, 160^\circ \right]$) or in the opposite narrow polar one ($\theta \in \left[ 0^\circ, 20^\circ \right]\cup\left[ 160^\circ, 180^\circ \right]$);
\item  particles are injected in a narrow equatorial sector ($\theta \in \left[ 70^\circ, 110^\circ \right]$) or in the opposite wide polar one ($\theta \in \left[ 0^\circ, 70^\circ \right]\cup\left[ 110^\circ, 180^\circ \right]$).
\end{enumerate}

\begin{figure}[h!]
\begin{center}
\includegraphics[width=0.5599999999999999\columnwidth]{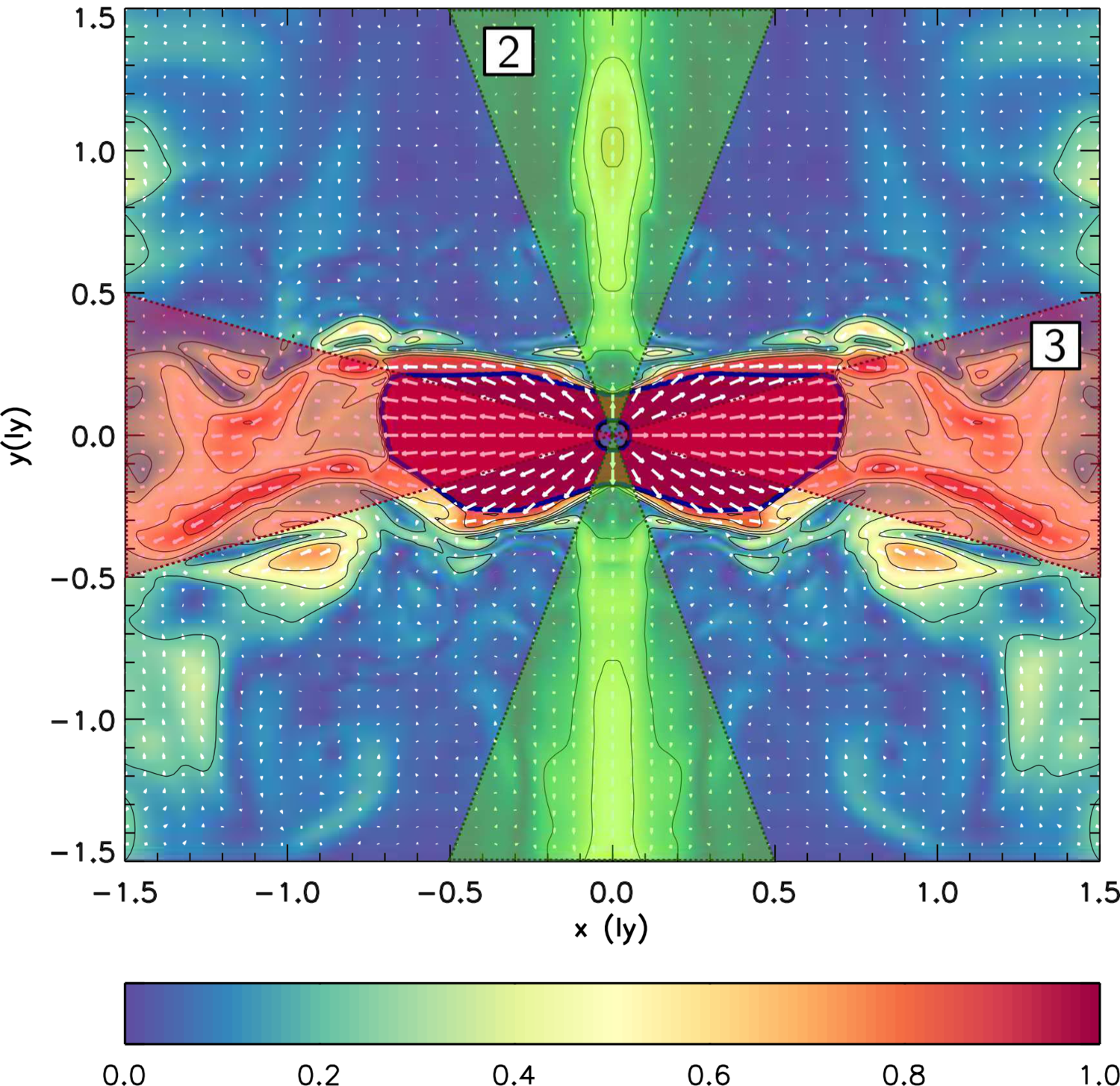}
\caption{Plot of the flow structure, with colors indicating the velocity magnitude in terms of $c$ and arrows the velocity field direction. The TS front is highlighted by the blue contour line of $0.99c$. Sectors corresponding to case (2) (polar) and case (3) (equatorial) are represented, respectively, by the green and red cones. \label{fig:ts}
}
\end{center}
\end{figure}

The polar and equatorial angular sectors of case (2) and case (3) are represented, respectively, by the green and red cones of Figure~\ref{fig:ts}, where the structure of the velocity field is shown in the proximity of the TS.

An explanation for the choice of these injection sectors is in order. The equatorial sector of case (3) roughly represents the wind striped region in the considered wind model (for a complete description see \citep{Olmi:2014}), while the polar sector of case (2) mimics a scenario in which particles can be accelerated in the low-$\sigma$ region around the polar axis thanks to Fermi I acceleration or O-point dissipation.

\subsection{Wisp profiles in the different scenarios}

The three hypotheses are tested for radio, optical and X-ray particle families.
In particular two different distribution functions are defined to account for radio and X-ray emission, while the optical one is obtained as a mixed contribution of the other two. 
Since radio and X-ray distribution functions are normalized in order to fit the complete integrated spectrum of Crab, the optical spectrum is naturally determined as the superposition of radio and X-ray contributions.

Radio particles are injected with the following spectrum:
\begin{equation}
f_{0\mathrm{R}}(\epsilon_0)\propto \left\{
\begin{array}{lcl}
0 & {\rm if} & \epsilon_0 < \epsilon_{\rm minR}, \\
\epsilon_0^{-p_\mathrm{R}} \exp(-\epsilon_0/\epsilon_\mathrm{R}^*) &  {\rm if} & \epsilon_0 > \epsilon_{\rm minR},\\
\end{array}
\right.
\end{equation}
and X-ray ones with:
\begin{equation}
 f_{0\mathrm{X}}(\epsilon_0)\propto \left\{ 
 \begin{array}{lcl}
 0 & {\rm if} & \epsilon_0 < \epsilon_{\rm minX}, \\
\epsilon_0^{-p_\mathrm{X}} \exp(-\epsilon_0/\epsilon_\mathrm{X}^*) &  {\rm if} & \epsilon_0 > \epsilon_{\rm minX}, \\
\end{array}
\right.
 \label{eq:Xpart}
\end{equation}
 where $\epsilon_0$ is the Lorentz factor of the particle at the injection site.
 
Power-law indices and cut-off energies that appear in the previous formulas, are all determined based on the comparison of the simulated emission with data. The best set of those parameters results to be: $p_\mathrm{R}=1.6$, $\epsilon_{\rm minR}=10^3$, $\epsilon_\mathrm{R}^*=2 \times10^6$, $p_\mathrm{X}=2.8$, $\epsilon_{\rm minX}=1.5 \times 10^6$ and $\epsilon_\mathrm{X}^*=10^{10}$.

\begin{figure}[h!]
\begin{center}
\includegraphics[width=0.84\columnwidth]{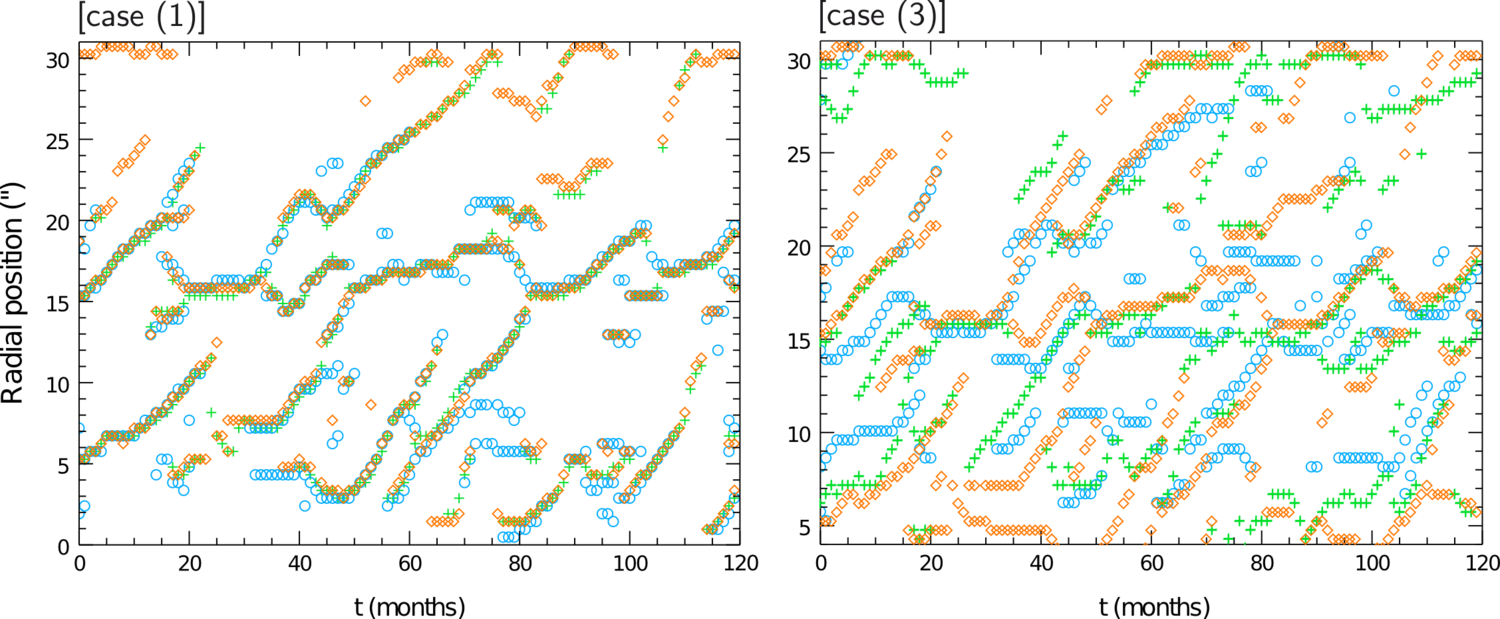}
\caption{Radial positions of the local intensity maxima (in arcseconds) as a function of time (in months) with orange diamonds identifying radio wisps ($\nu_r=5$ GHz), green crosses optical ones ($\nu_o=3.75 \times 10^{14}$ Hz) and light-blue circles for X-rays (1 keV). 
On the left case (1) is shown. On the right case (3) is shown, with X-ray particles injected in the equatorial zone and radio ones injected in the complementary sector. 
\label{fig:wisp1}
}
\end{center}
\end{figure}

The first interesting point to look at is the difference between wisps profile under case (1) assumptions (i.e. uniform injection at TS) and one of the other two cases. In Figure~\ref{fig:wisp1} wisp profiles are shown as plots of the radial position of the intensity maxima as a function of time, with case (1) on the left and case (3) (X-ray particles injected in the equatorial sector and radio in the polar one) on the right.
As expected, wisps appear to be coincident at the different wavelengths in the hypothesis of case (1). 
This is no more true as soon as particles of different families are injected in different sectors of the shock, as can be easily seen in the plot on the right.

\begin{figure}[h!]
\begin{center}
\includegraphics[width=0.84\columnwidth]{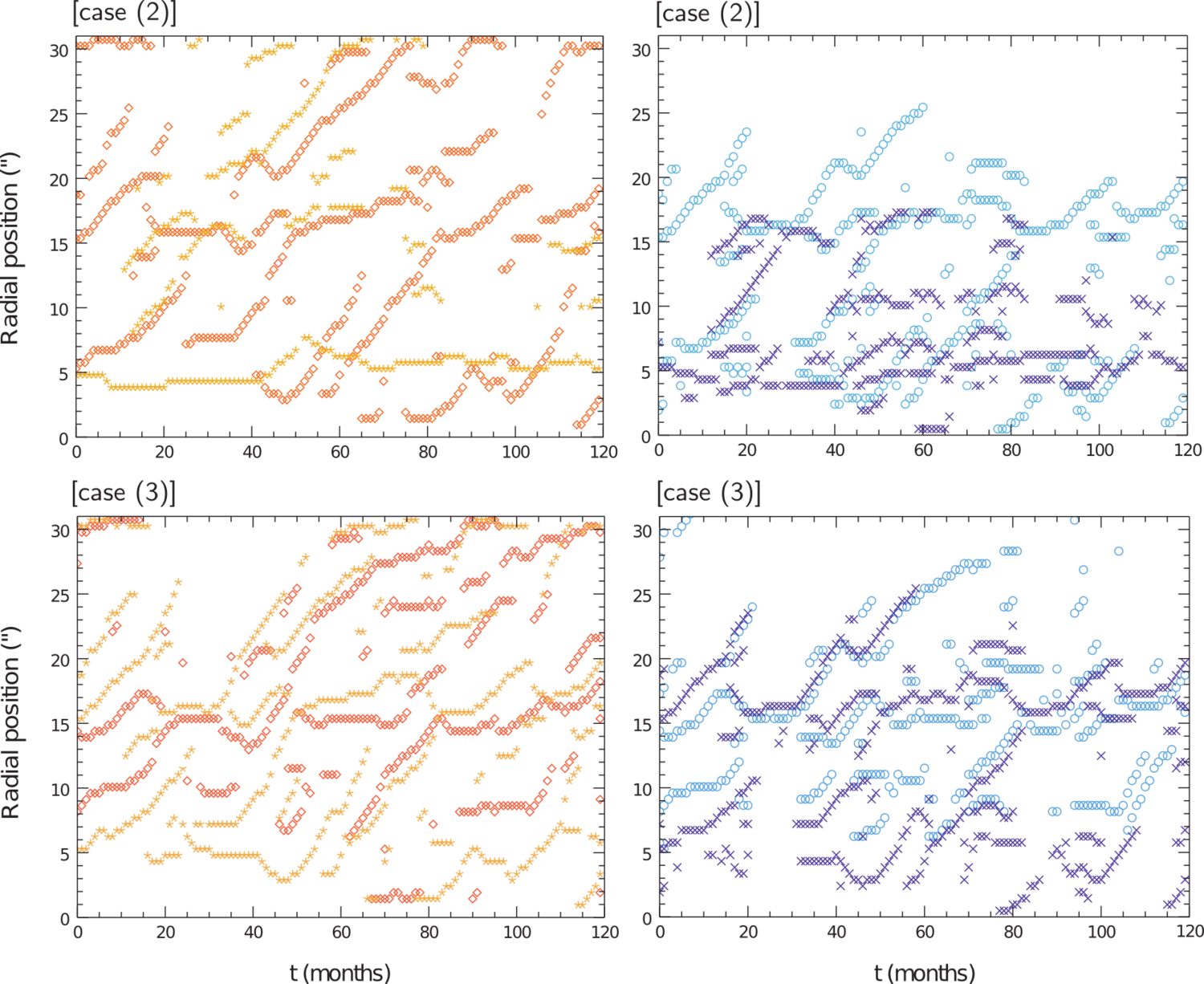}
\caption{Radial position of the local intensity maxima as a function of time for case case (2), in the upper row, and case (3), in the bottom one. On the left, different possibilities for the injection of radio particles are shown. Stars and diamonds represent, respectively, polar and equatorial injection. On the left-hand side, the same cases are shown for X-ray particles: violet crosses are used for polar injection and light blue circles for equatorial one. 
\label{fig:wisp2}
}
\end{center}
\end{figure}

Excluding the case of uniform injection, which clearly does not reproduce the expected behavior of wisps, the remaining  possibilities are case (2) and (3). 
All the different scenarios of these two cases are shown and compared in Figure~\ref{fig:wisp2}. 
Here particles responsible for radio and X-ray wisps are injected either in the polar sector (stars and crosses) or in the equatorial region (diamonds and circles).

When particles are injected in the wide equatorial region of case (2), wisps appear to be almost identical of those of the uniform injection case. 
On the contrary, particles injected in the narrow polar sector are not seen to produce wisp-like features, with the most prominent structure being a quasi-stationary feature at a distance of $5^{\prime\prime}$ from the pulsar, which is something very different from what observations show.

Under hypotheses of case (2) it is thus impossible to define two non coincident sectors in which radio and X-ray particles can be injected, so as to give rise to different wisps at the different energies.

Finally, in the bottom row of the same Figure~\ref{fig:wisp2}, the alternative injection scenarios considered in case (3) are shown. 
Here wisps appear in any case, no matter whether particles are injected in the wide polar sector or they are injected in the narrow equatorial region. 
The strongest constraint for discriminating between the possibilities comes from \cite{Schweizer:2013}, 
where X-ray wisps were shown to be present only beyond $\sim 6^{\prime\prime}$ from the pulsar. 
The only case in which this behavior is correctly reproduced is when X-ray particles are injected in the equatorial region, which approximatively corresponds to the striped zone of the wind, where dissipation of magnetic field is most efficient.
This may indicate that X-ray particles are produced via Fermi I acceleration in the striped zone, where magnetization can be low enough to make this mechanism viable.

Outward apparent velocities of wisps at the different frequencies have also been investigated for case (3), leading to a range of $0.08c \lesssim v \lesssim 0.38c$, which is in good agreement with what observed.

Strong constraints on radio emission are on the other hand very difficult to draw, and the only case that can be effectively excluded is the one in which radio particles are injected in a narrow polar cone. 
But, in order to have non coincident wisps at the different wavelengths, radio particles must be accelerated elsewhere than the equatorial region of case (3). 
The remaining possibilities are thus that acceleration happens in a wider equatorial region or in the complementary polar zone with respect to X-ray particles, where conditions for driven magnetic reconnection to be at work might be locally satisfied.

\section{Gamma-ray flares of the Crab nebula}
\label{sec:flares}

\subsection{The flare observations}

\begin{figure}[h!]
\begin{center}
\includegraphics[width=0.7\columnwidth]{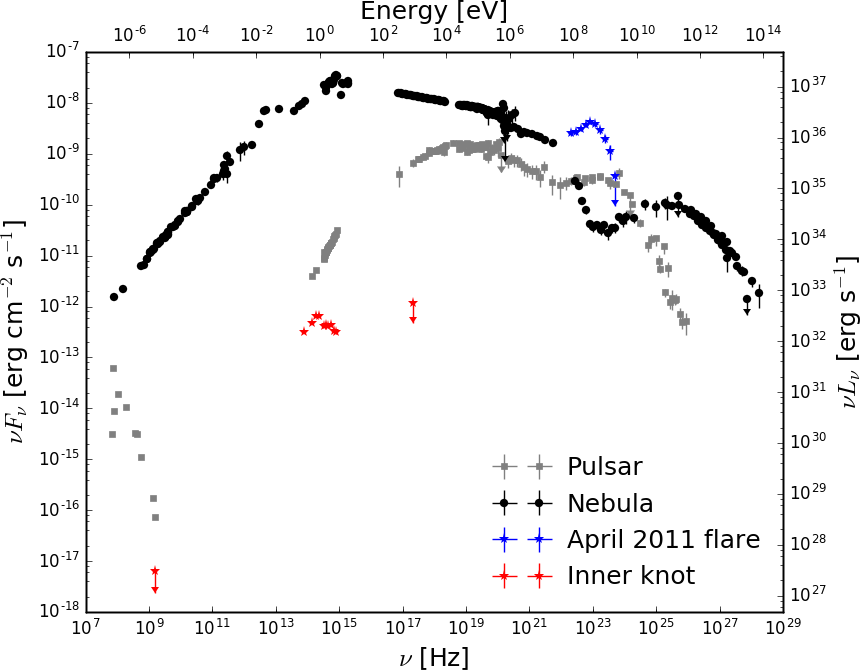}
\caption{SED of the Crab nebula and pulsar. See \citet{BuehlerBlandford2013a} for the references to the data from the average nebula and pulsar SEDs. The Crab knot spectrum is also shown, its data was was taken from \citet{Lobanov2011,Sandberg2009,rudy-15}. Also shown is the SED at the maximum of the April 2011 gamma-ray flare from \citet{buehler2012}.\label{fig:sed}
}
\end{center}
\end{figure}

\begin{figure}[h!]
\begin{center}
\includegraphics[width=1\columnwidth]{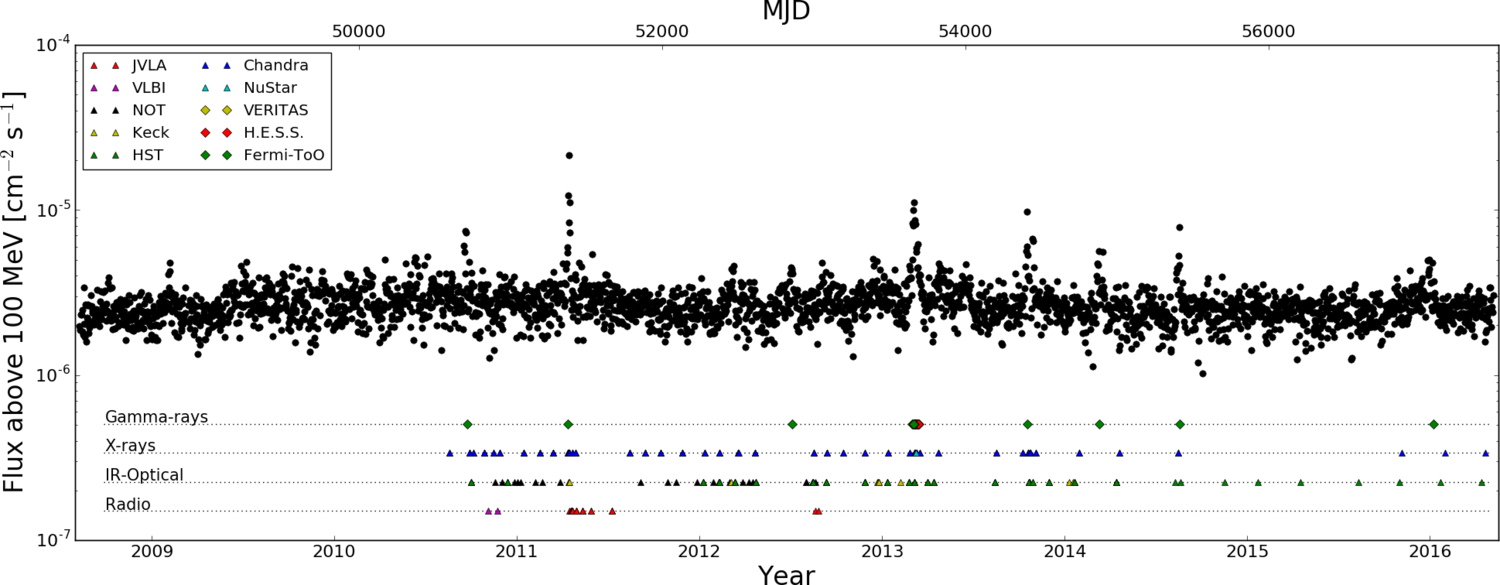}
\caption{The black markers show the daily averaged gamma-ray flux above 100 MeV from the direction of the Crab nebula measured by the Fermi-LAT. This data is routinely updated and publicly available at \url{http://fermi.gsfc.nasa.gov/ssc/data/access/lat/msl_lc/}. Colored markers indicate the times at which instruments at other wavebands observed the source (see text).\label{fig:lc}
}
\end{center}
\end{figure}

Particle acceleration was observed in real time for the first time in the Crab nebula with the discovery of the gamma-ray flares in 2010.  The nebula emission was expected to  be stable over time scales of years. It therefore came as a huge surprise when the AGILE and Fermi-LAT satellites observed strong high energy (HE, $>\rm  100 MeV$) gamma-ray flares \citep{Tavani2011,Abdo2011}. When observed in gamma-ray band with energies above 100 MeV, the nebula emission is highly variable. In this energy band one observes the high energy end of the synchrotron component, and the onset of the inverse Compton component in the SED of the nebula, as shown in Figure \ref{fig:sed}. The monthly flux variations observed in this frequency band in the first 8 years of the Fermi-LAT mission are shown in Figure \ref{fig:lc}). A statistical analysis of the flux variations shows that the nebula emission in HE gamma-rays varies on all time scales that can be resolved by current instruments. The power density distribution (PDS) of the frequency of flux variation can be described by a red noise process with an index of $PDS \propto \nu^{-0.9}$  from yearly to daily time scales \citep{buehler2012}.

Extreme gamma-ray outbursts are observed approximately once per year. During these flares the photon flux above 100 MeV can increase up to a factor $\approx 30$. During the two brightest flares in April 2011 and March 2013 flux doubling time scales of $t_{var}\approx 6$~ hours were observed \citep{buehler2012,Mayer2013}. The isotropic luminosity release at the peak of these flares was $L_{\gamma, iso} \approx 2 \times 10^{36}$~erg s$^{-1}$, about 1\% of the spin-down power of the pulsar. The temporal structure of the flares is generally different between them, as are their spectral properties. Their photon flux, $f_{ph}$,  as a function of energy is typically well described by a power law $f_{ph} \propto E^{-\gamma}$, with  wide range of photon indices ranging between  $\gamma \approx 1.2 - 3.7$. For the flares of April 2011 and March 2013, a high energy cutoff of the spectrum could be statistically resolved (see Figure \ref{fig:sed}). The cutoff energy is approx $E_{cut} \lesssim 1$ GeV. This appears to be a general feature of the flare spectra, for no significant emission was detected beyond this energy. Interestingly, the time evolution of the cutoff could be resolved in time during the April 2011 flare. It was observed that the total energy flux of the gamma-ray emission, $f_e$,  scales as a function of the cutoff energy as $f_e \propto E_{cut}^{3.4 \pm 0.9}$  \citep{buehler2012}.

Unfortunately, the angular resolution in gamma-rays is not sufficient to pinpoint the site of the emission region within the nebula. However, observations at lower frequencies can resolve the nebula morphology in great detail. Therefore, extensive multi-wavelength campaigns have been in place since 2010 to find correlated emission to the gamma-ray flares at lower frequencies. Some of the most sensitive instruments in the world participated in these multi-wavelength campaigns, as the Hubble Space Telescope (HST), the Keck Observatory, Chandra X-ray Observatory \citep{Weisskopf2013,rudy-15}, the Very Long Baseline Interferometer \citep{Lobanov2011}  the Jansky Very Large Array \citep{Bietenholz2014}, the H.E.S.S., VERITAS and MAGIC Cherenkov telescopes \citep{Abramowski2014,Aliu2014,Aleksic2015} and NuSTAR \citep{Madsen2015}. The dates of these observations are shown in Figure~\ref{fig:lc}. In total more than 1 Msec of observations were taken by these  instruments since 2010. Simultaneous observations on monthly intervals were carried out and additional observations were taken in intervals of a few days during times in which the HE gamma-ray flux was high. Surprisingly, to date, these observations have not revealed any emission correlated to the gamma-ray flares \citep{Weisskopf2013,rudy-15}.

\subsection{Theoretical models of the flares}

The Crab nebula flares pose severe challenges to models of particle acceleration. It is usually assumed that the flare emission originates from synchrotron radiation of freshly accelerated electrons. The reason is that the high gamma-ray energy of the flares ($\approx 1$~GeV) implies the presence of electrons with multi PeV energies for typical magnetic fields of $B \approx 200 \mu$G expected in the nebula. The cooling time of these particles is $\tau_{cool} \lesssim 20$ days. Therefore the emission region cannot be far from the acceleration site  (however, other views have also been proposed, see e.g. \citet{Bykov_2012} and \cite{2015arXiv151205426Z}). Assuming the acceleration and radiation region are the same, the high luminosity of the flares poses severe constraints on the acceleration efficiency during these events: the total isotropic fluence during the brightest flares is $\varepsilon \approx L_{\gamma,iso} \times t_{var} \approx 4 \times 10^{40} $~erg (again adopting $t_{var}\approx 6$~hours). Causality implies that the emission region has a volume $\approx (c t_{var})^3$ if there is no strong Doppler beaming. The magnetic energy in such a region is $\varepsilon_B \approx \frac{c^3}{8 \pi} t_{var}^3 B^2$. The radiation efficiency can then be estimated as $\epsilon \equiv \varepsilon / \varepsilon_B \approx  3700 \times B_{-3}^{-2}$, where $ B_{-3}$ is the magnetic field measures in mG.

The RMHD simulations discussed in Section \ref{sec:rmhd} show that magnetic fields reach a maximum of a few mG within the body of the nebula. This implies a radiation efficiency $\epsilon \gg 1$ for the flare emission. This contradiction shows that flare emission cannot be isotropic and that relativistic beaming is likely playing an important role. However, it is not easy to explain highly relativistic plasma motions in the nebula: these are not found in RMHD simulations and observations show only mildly relativistic speeds, of $\approx$ 0.5c on the resolvable spatial scales. It is therefore likely, that a combination of boosting and a high magnetic field in the emission region is required. Furthermore, the field energy in this macroscopic region must be transferred very rapidly to kinetic particle energy (macroscopic compared to the kinetic scales as the plasma skin depth). 

The general concept of catastrophic dissipation of magnetic energy to non-thermal particles on macroscopic scales has been named \textit{magnetoluminescence} by \citet{Blandford_2014}. The process might be triggered by ideal MHD instabilities on large scales and produce regions with non-ideal conditions where rapid particle acceleration takes place. Recently, several works have studied this concept using force-free/MHD \citep{East_2015,LyutikovSironi2016,Zrake_2016} and PIC simulations \citep{LyutikovSironi2016,Nalewajko_2016,Yuan_2016}. For example, \citet{East_2015} found that highly magnetized plasma configurations, in the form of multiply packed flux ropes in a 3D periodic box, can be unstable to ideal modes if they possess "free energy"--the amount of energy dispensable while preserving the overall topology. These instabilities evolve rapidly and release the magnetic free energy over a single dynamic time scale.  PIC simulations further showed that the large scale instability forces current sheets to form self-consistently over dynamic time scales---these are the main sites of particle acceleration and electromagnetic dissipation \citep{LyutikovSironi2016,Nalewajko_2016,Yuan_2016}. Similar to the planar current sheet reconnection case \citep{Cerutti_2013,Cerutti_2014}, particles accelerated in the current sheets are beamed and at the same time bunched by the tearing modes, resulting in rapid variability of observed synchrotron radiation \citep{Yuan_2016}. While it is unlikely that these highly stylized flux rope configurations used in simulations correspond to realistic field structures in the Crab nebula, they provide some insights into rapid dissipation of electromagnetic energy. \citet{LyutikovSironi2016} also studied similar processes in a more natural configuration, where two adjacent flux tubes with a zero total current merge and accelerate particles efficiently. It still remains to be seen if the aforementioned schemes produce all the features of the Crab flares, but the progress so far makes them promising candidates.

Independent of the actual particle acceleration process, it appears clear from the discussion above that regions of high magnetization and large relativistic motion are preferred sites for the gamma-ray flares. The intermediate latitude region just downstream of the reverse shock is one of the most promising regions \citep{LyutikovSironi2016}. Interestingly, the shock in this region might correspond to the inner knot, observed very close to the pulsar, as will be discussed in Section \ref{sec:knot}. 

\subsection{Do the flares originate from the inner knot?} 

\begin{figure}[htbp]
\begin{center}
\includegraphics[width=0.7\columnwidth]{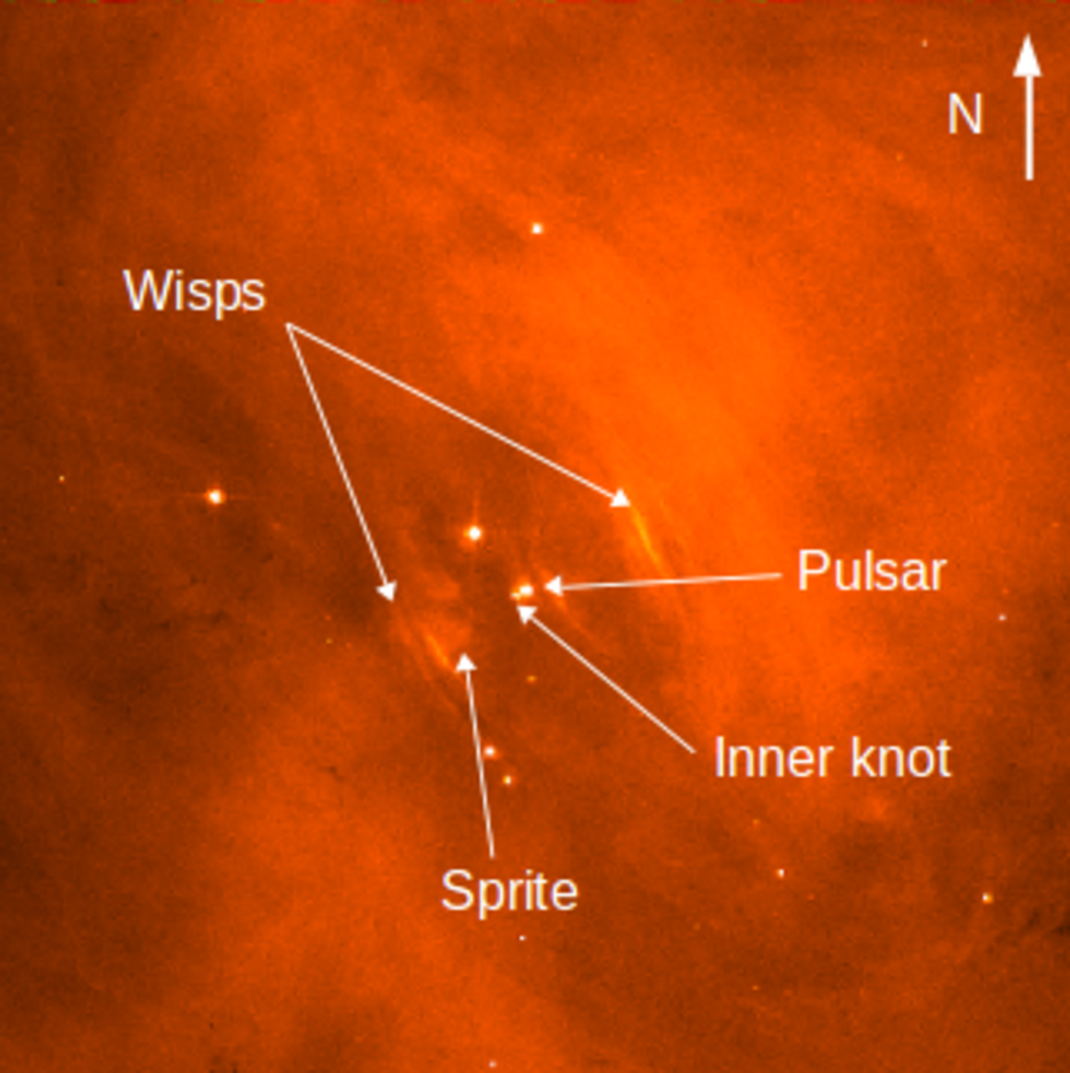}
\caption{\label{fig:knot}
Optical image of the Crab nebula taken by the Hubble Space Telescope. The image is 0.02$^\circ$ on each side. Different structures of the nebula are indicated \citep[e.g.][]{hester2002}. Particularly, the ``inner knot'' can be seen, it is located $\approx 0.6$ arc sec away from the pulsar.
}
\end{center}
\end{figure}

The inner knot is the closest feature observed around the pulsar, located $\approx$0.6 arc sec south east of it (see Figure \ref{fig:knot} ). The knot is detected in the infrared and optical bands. In radio and X-rays it has so far not be detected, the obtained upper limit on the flux are shown in Figure \ref{fig:sed}. The inner knot is a dynamical feature, its distance to the pulsar can vary by $\approx$20\%  and its brightness by $\approx$50\% \citep{Sandberg2009}.  Its emission has a high degree of linear polarization \citep{Moran_2013}. Recent observations suggest that the polarization angle might change over time \citep{Moran_2015}.

\begin{figure}[htbp]
\begin{center}
\includegraphics[width=0.7\columnwidth]{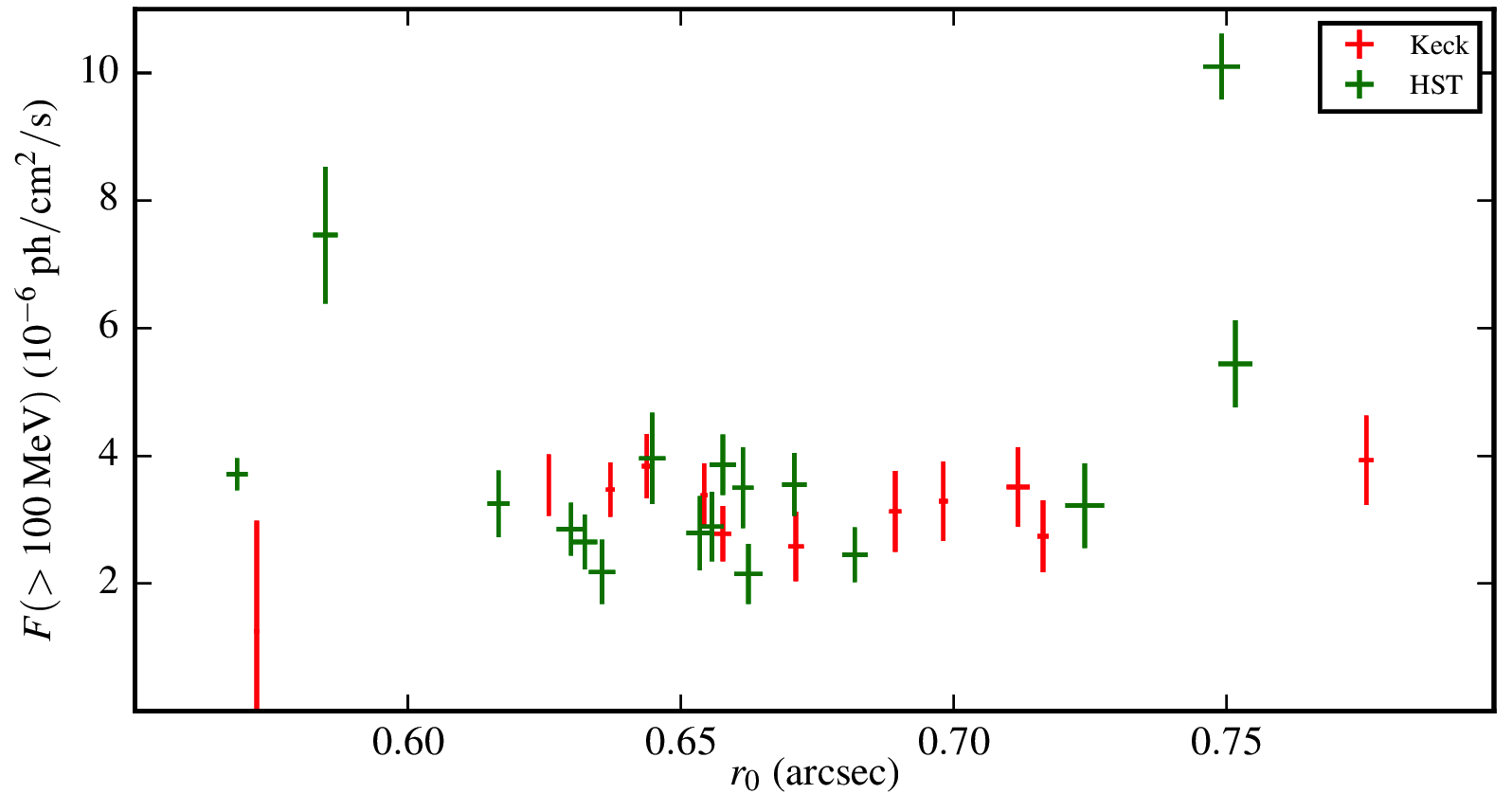}
\caption{\label{fig:knotgamma}
Gamma-ray flux above 100 MeV as a function of the distance of the inner knot to the Crab pulsar. The figure is reproduced from \citet{rudy-15}.  
}
\end{center}
\end{figure}

Its variability, compactness, high-level of polarization and proximity to the pulsar made the inner knot one of the prime candidates for the site of origin of the gamma-ray flares.  In fact, it was proposed by \citet{ssk-lyut-11} that almost all of the emission above 100 MeV could come from this feature. Therefore, particularly dense observations targeting the inner knot where performed by the Keck Observatory and HST. Surprisingly however, none of the knot's properties was found to correlate with the gamma-ray emission \citep{rudy-15}. One example is shown in Figure \ref{fig:knotgamma},  where the gamma-ray flux is plotted as a function of the distance of the knot from the pulsar. The inner knot is therefore likely not the site of origin of the gamma-ray flares.
\begin{figure}[htbp]
\begin{center}
\includegraphics[width=1\columnwidth]{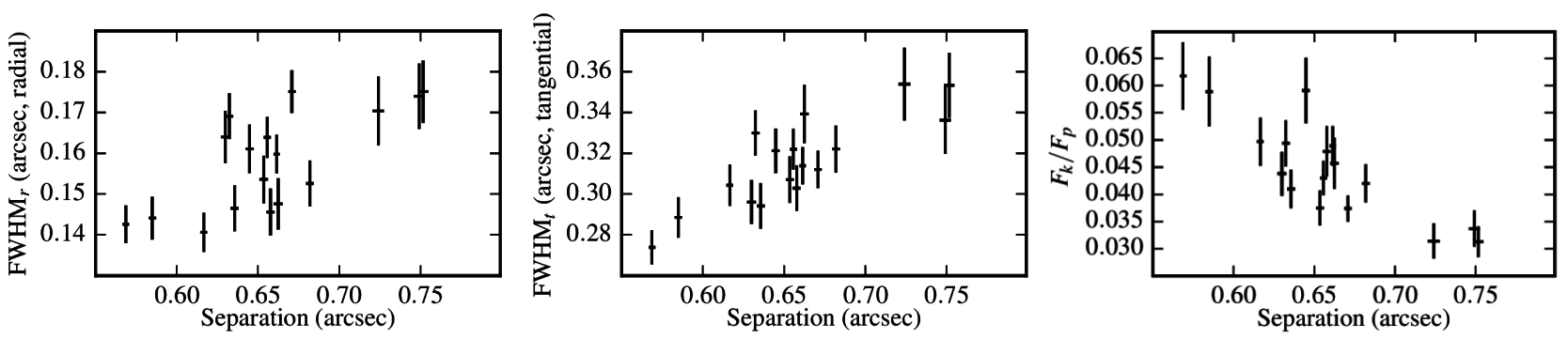}
\caption{\label{fig:knotcorr}
Properties of the inner knot as a function of its distance from the Crab pulsar. \textbf{Left:} size of the knot perpendicular to the pulsar-knot axis, \textbf{Middle:} size of the knot along the pulsar-knot axis, \textbf{Right:} k-band flux of the knot normalized to the pulsar flux. The figures are reproduced from \citet{rudy-15}.
}
\end{center}
\end{figure}
However, the observations revealed the dynamical behaviour of the inner knot with unprecedented detail. Its size was found to increase with the distance of the pulsar. At the same time its luminosity decreases with increasing distance, as shown in Figure \ref{fig:knotcorr}.  As will be presented in detail in Section \ref{sec:knot}, modelling of the knot shows that it is likely the high-latitude shock of the pulsar wind, giving us the first radiative signature of the wind particles.

\section{Inner knot of the Crab Nebula}
\label{sec:knot}

The discovery of the jet-torus structure of the inner Crab nebula coincided with the emergence of computer codes for relativistic MHD capable of dealing with highly relativistic magnetised flows \citep{ssk-godun99}.  This was most handy, as theoretical MHD models of the Crab nebula allowed for reasonable ideas on the origin of the structure but could not address the problem rigorously due to the complicated nature of non-spherically symmetric flows.   The  key properties of pulsar winds giving rise to the jet-torus 
appearance of the nebula in these models  are 1) their anisotropy, with the wind power increasing towards the equatorial plane of the pulsar rotation 2) their magnetisation, with purely azimuthal magnetic fields aligned with the pulsar's rotational axis.  The first property naturally leads to the torus component, whereas the second opens the possibility of magnetic collimation of the flow toward the polar axis. In fact, this collimation  mechanism fails for the wind itself, with possible exception for a tiny polar section, due to its highly super-magnetosonic motion. However, downstream of the wind termination shock the corresponding Mach number drops, the causal connectivity across the flow is restored and the magnetic hoop stress regains its collimating  potential.  Thus, in the MHD model the Crab's jet is not present in the un-shocked pulsar wind but forms in the shocked part of the wind and thus already inside the nebula.   

The first computer-generated models of the Crab nebula focused on the case of weakly magnetised pulsar wind, thus adopting the conclusions of the Rees-Gunn-Kennel-Coroniti theory \citep{rees-gunn-74,kc84a,kc84b}.  To the delight of theorists, the simulations confirmed the possibility of the separation of the post-shock flow into the equatorial (torus)  and polar (jets) components, provided the wind magnetisation parameter $\sigma$ exceeded the critical value of few $\,\times\, 10^{-3}$ \citep{ssk-lyub-03,ssk-lyub-04,del-zanna2004,bogovalov2005}.  This could be concluded immediately from the analysis of the velocity field and the distribution of other fluid parameters but in order to compare the results with the observations a synthetic emission map is highly desirable. Some elements required for the synchrotron emission calculations were readily provided by the numerical models. These are the magnetic field strength and the orientation, as well as the velocity field, which is important for the relativistic beaming and Doppler effects.   The missing part, concerning the spectrum of the ultra-relativistic electrons, had to be recovered indirectly, via a crude model connecting the spectrum to the fluid pressure of the MHD solution\footnote{A more sophisticated technique was used in more recent simulations. }.  The un-shocked wind zone was not expected to make a significant contribution and hence was explicitly excluded from the emission calculations.

\begin{figure}[h!]
\begin{center}
\includegraphics[width=0.7\columnwidth]{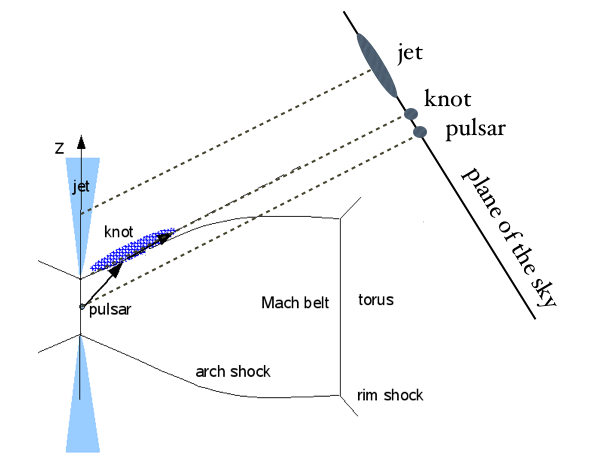}
\caption{\label{knot-mhd-model}
In this sketch, we show the plane defined by the rotational axis of the pulsar and the location of fiducial  observer.  The region around the pulsar corresponds to the un-shocked  pulsar wind, the continuous lines show the wind termination shock, indicating its topological components, the Mach belt, arch and rim shocks. The blue shaded region shows the region where the shocked wind plasma flows towards the observer with moderately relativistic speed. The dashed lines are the lines of sight, which show the order in which the jet the knot and the pulsar appear in the plane of the sky of the observer.
}
\end{center}
\end{figure}

One of the most prominent and yet unexpected features of the synthetic synchrotron images of the simulated PWN was a very bright feature located right in the centre, where the image of the model's pulsar would be if its emission were included (but it was not: \citet{ssk-lyub-03,ssk-lyub-04}).  A closer look revealed that the feature was a slightly off-centred and extended knot.  In order to identify the feature of the synthetic images with a particular feature of the numerical solution,  a detailed inspection of the data was carried out and it revealed that the emission originated close to the termination shock, where the shocked wind plasma was still flowing with  (moderately) relativistic speed towards the fiducial observer,  and for this reason was subject to significant Doppler-boosting.  This is illustrated in Figure~\ref{knot-mhd-model}. Subsequent  studies by other groups \citep{delzanna-06} and  the recent 3D simulations \citep{PorthKomissarov2013,PorthKomissarov2014a} confirmed the result.    

Once we had understood the nature of the feature, it became clear that a counterpart must be present in the real images of the Crab nebula, unless something is very wrong with the MHD model. The oblate geometry of the termination shock ensures that there always be a shock Section where the post-shock plasma is streaming towards the observer with relativistic speed and hence its emission is Doppler-boosted.   Provided the Crab's synchrotron electrons are indeed accelerated at the termination shock, as proposed in the Kennel-Coroniti model, this ensures the existence of a bright knot inside the area occupied in the plane of the sky by the un-shocked pulsar wind zone.  This knot must be a very bright permanent feature, located on the projected symmetry axis. Since the polar jets are also subject to relativistic beaming, the knot must be on the jet side of the pulsar (see Figure~\ref{knot-mhd-model}) and have no counterpart on the counter-jet side.      
A comparison with the high-resolution HST images revealed that such a feature is indeed present  in the Crab nebula --
it is called the inner knot (or knot 1; \citet{hester-95}) and is located approximately 0.65 arcsec away from the Crab pulsar (see Figure~\ref{fig:knot}).  
So far it has been detected only in the optical-IR range (the highest energy detection was in NUV band by \citet{Melatos_2005}).  The emission is strongly polarised with the electric polarisation vector aligned with the rotational axis in the plane of the sky. This is exactly what is expected, as in the close vicinity of the termination shock the magnetic field should still preserve the highly regular azimuthal structure it has in the wind.  In a way, the inner knot was predicted by the simulations and then confirmed by the observations. Indeed, even if the observational discovery of the Crab's knot  preceded  the simulations, it was not immediately connected to the termination shock and its emergence in the synthetic maps was a complete surprise.      

The identification of the inner knot with the termination shock is already very interesting as it implies a unique opportunity for studying properties of highly relativistic shocks in magnetised plasma.  The discovery of gamma-ray flares in the Crab nebula added interest as the short life-time of the electrons emitting at such energies suggests that the gamma-ray emission originates from the termination shock, provided it is the main acceleration site for the synchrotron electrons of all energies. Moreover, the beamed nature of the emission in this region ensures the domination of  the inner knot contribution to the observed 
gamma-ray flux \citep{ssk-lyut-11}.  Unfortunately, the angular resolution of gamma-ray telescopes is not sufficient to test this conclusion directly.  Under such circumstances, the only way to localise the flares is via their identification with structural variability in the radio, optical-IR and X-ray windows where the resolution of modern instruments is much higher. Such observations have been carried out but they have not led to a positive identification so far. Moreover, they seem to have ruled out the inner knot as the source of gamma-ray flares as its variability did not show any correlation with the flares \citep{rudy-15}. On the other hand, these studies significantly  increased the available information on the knot properties, allowing to test its theoretical connection with termination shock in greater detail.   

The knot does not seem to have much of an internal structure, with brightness gradually decreasing from the peak value in the centre to the background one at the periphery. It is elongated in the direction normal to the rotational axis and a little bit bowed away from the pulsar. The knot is clearly separated from the pulsar and the separation varies with  a $30\%$ amplitude on the time scale of few months, which is similar to the wisp production time scale.   The knot's flux anti-correlates with the separation.      
The variability implies that in the vicinity of the termination shock the flow of shocked plasma is not steady but varies on the time-scale of the shock light-crossing time.  In fact, such a variability, accompanied by a strong variation of the shock geometry, was  discovered in numerical simulations before the observations. In the high-resolution 2D simulations by \citet{camus2009}, it was found that this variability led naturally to emergence of wisps in synthetic synchrotron maps.  These wisps were identified with inhomogeneities created by the variable shock in the  equatorial outflow and advected with the outflow speed.  Thus, the similar time scales of the inner knot dynamics and the wisp production are easily explained by the fact that both are traced back to the termination shock variability.  Moreover, the recent 3D simulations of the Crab nebula are consistent with the observed knot's flux-separation anticorrelation \citep{PorthKomissarov2014a}.    

\begin{figure}[h!]
\begin{center}
\includegraphics[width=0.7\columnwidth]{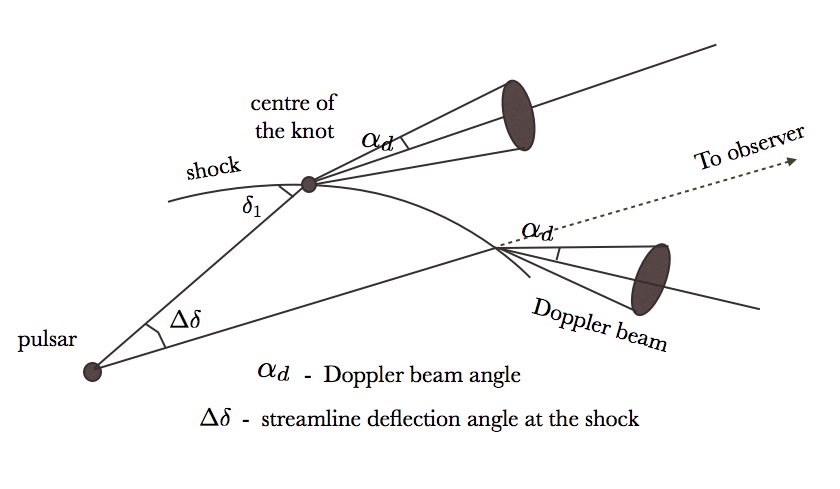}
\caption{\label{knot-separation}  {In this sketch, we show Doppler beams at two locations on the termination shock, selected by the lines of sight originated at the pulsar and  at the brightness peak of the knot, where the post-shock velocity is parallel to the line of sight.  
In order for the knot not to extend all the way to the pulsar in the plane of the sky,  the Doppler beam at the first location should point away from the observer and hence $\alpha_d < \Delta\delta$.  
}
}
\end{center}
\end{figure}

A couple of attempts have been made recently to build a simple semi-analytical model of Crab's inner knot \citep{YB-15,LyutikovKomissarov2016}. The main motivation behind these attempts is based on the fact that the structure of the upstream wind is relatively simple and one of the key factors determining the knot appearance, namely the Doppler beaming, is relatively easy to predict based on the properties of relativistic transverse MHD shock.  The most important conclusion of the studies is that the observed clear separation of the knot from the pulsar can only be achieved in models with low wind magnetisation ($\sigma \ll 1$).  Indeed, suppose that the brightness peak of the knot corresponds to the point on the shock where the shocked plasma flows directly towards the observer. Then the observed angular distance between the peak and the pulsar is determined by flow deflection angle $\Delta\delta$ at the shock (see Figure~\ref{knot-separation}). The knot will not engulf the pulsar provided the half-opening angle of the Doppler beam, $\alpha_d$, is smaller than the deflection angle along the line of sight passing through the pulsar (see Figure~\ref{knot-separation}). Given the ultra-relativistic nature of the wind, both these angles depend mostly on the upstream $\sigma$ and hence the condition $\alpha_d < \Delta\delta$ translates into the condition on $\sigma$.    

The deduced low $\sigma$ is in contrast with the theoretical expectation of high $\sigma$ in pulsar winds. This conflict can be settled if the knot is produced by the striped equatorial component of the wind. In the striped section, $\sigma$ can be reduced via
magnetic dissipation of the stripes either in the wind or inside the hybrid MHD shock which terminates it, with the post-shock flow dynamics being identical in both these cases.

\begin{figure}[h!]
\begin{center}
\includegraphics[width=0.7\columnwidth]{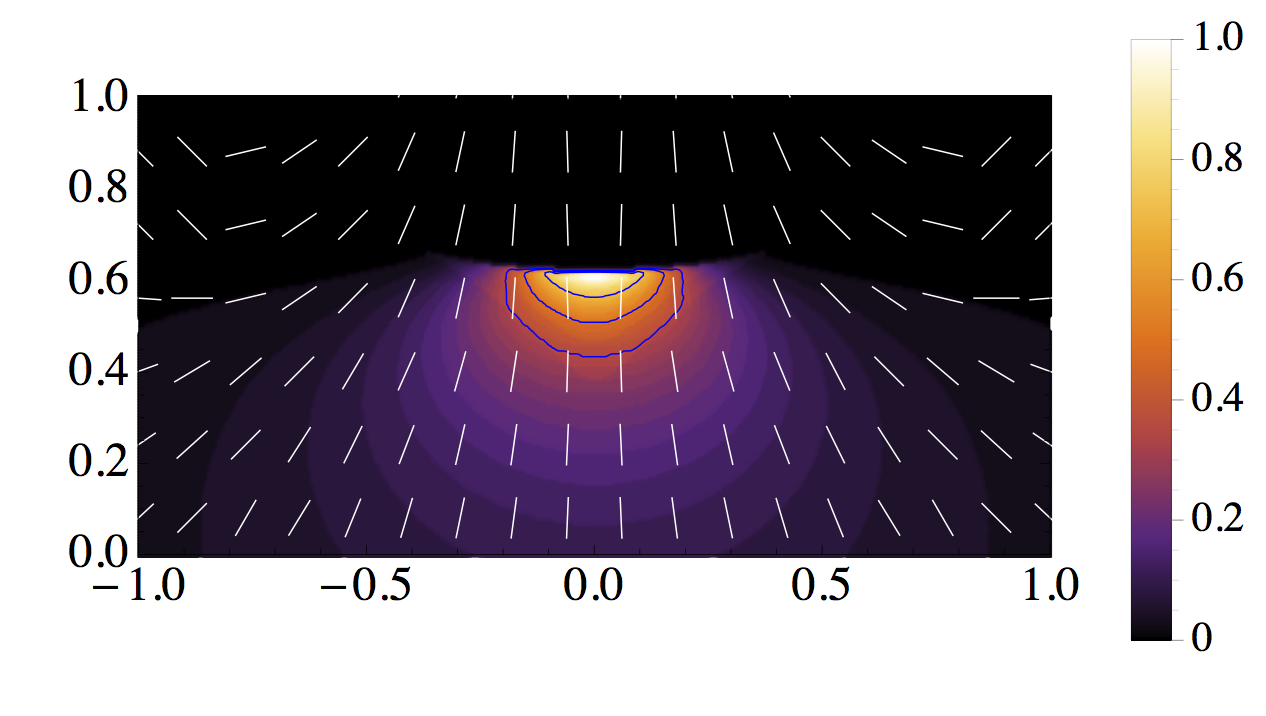}
\caption{\label{knot-polarisation}{The emissivity distribution over the shock surface projected onto the plane of the sky and the polarisation vector of the shock emission. 
}
}
\end{center}
\end{figure}

Using the Doppler-beamed post-shock emissivity as a proxy for the knot brightness, one can estimate a number of observed  knot parameters such as its elongation, the distance from the pulsar in the units of the equatorial radius of the termination shock, and its polarisation degree.     The results are somewhat dependent on the  utilised model for the termination shock shape, but generally agree with the observational measurements quite well provided the knot is formed by the low-sigma part of the wind.  
The total flux polarisation degree is strongly effected by the relativistic aberration of light, which leads to a noticeable rotation of the polarisation vector across the knot (see Figure~\ref{knot-polarisation}).  \citet{YB-15} concluded that the rotation imposes an upper limit 
of $\simeq 50\%$ on the overall polarisation degree of the knot,  which is significantly lower than the observed $\simeq 60\%$ \citep{Moran_2013}.  However, in the observations the integration was carried out only over the central part of the knot, thus excluding outer regions where the rotation is strong. \citet{LyutikovKomissarov2016} have demonstrated that the polarisation degree decreases with the size of the integration area and this can explain the disagreement  between the observations and the upper limit found by \citet{YB-15}.      
Moreover, the polarisation degree of the synthetic knot reproduced in the 3D simulation \citep{PorthKomissarov2014a} agrees with the observed values exceptionally well. 

\begin{figure}[h!]
\begin{center}
\includegraphics[width=0.9\columnwidth]{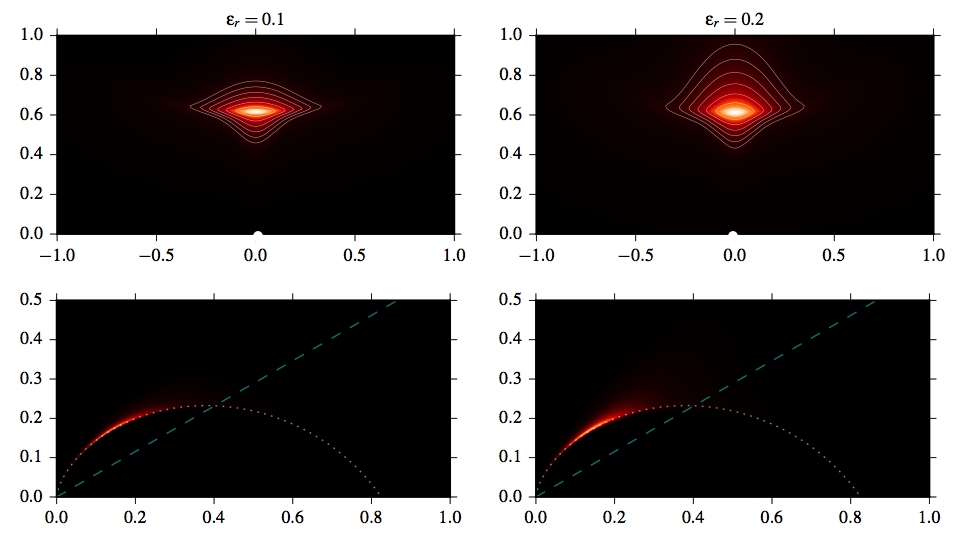}
\caption{\label{thickness-effect}{The top panels show the knot images resulting from emitting region of thickness equal to 0.1 (left) and 0.2 (right) of the local shock radius. The bottom panels illustrate the emissivity models used for these calculations. The dotted line shows the termination shock surface, the dashed line shows the line of sight and the mono-coloured image shows the corresponding volume emissivity distribution. 
}
}
\end{center}
\end{figure}

The shape of the knot based on the emissivity isophotes was another matter of concern \citep{YB-15}. Instead of being bowed away from the pulsar the knot looked bowed towards it, reflecting the shape of the termination shock ``shadow'' in the plane of the sky (see Figure~\ref{knot-polarisation}). In order to understand the significance of this result one has to study the role of other factors influencing the knot appearance. For example,  the effective geometric thickness of the emitting layer may have a strong impact as well as the variations of the velocity field in this layer \citep{LyutikovKomissarov2016,YB-15}.  Figure~\ref{thickness-effect} illustrates the role of finite thickness in a model where the emissivity above the termination shock is set via an extrapolation of that at the shock surface \citep{LyutikovKomissarov2016}. One can see that the effect is rather strong.

In these calculations, it was assumed that the spectrum of the knot emission is a power law. This implies that the termination shock is an efficient accelerator of non-thermal particles. Although this assumption is a corner stone of the popular and successful model for the Crab nebula emission by \citet{kc84b}, its validity has been questioned by the PIC simulations of shocks in striped flow \citep{Sironi:2011}. These suggest that the non-thermal particle acceleration can only be efficient in a small equatorial sector where the downstream plasma becomes almost unmagnetised.  Otherwise, the plasma is simply heated and its synchrotron emission has a thermal spectrum.  Interestingly, the current failure to detect both the radio and X-ray emission from the knot  is consistent with such a spectrum.      

\section{Gamma-ray binaries: pulsar winds interacting with a massive companion}
\label{sec:binaries}
\subsection{Introduction}
\begin{figure}[h!]
\begin{center}
\includegraphics[width=0.7\columnwidth]{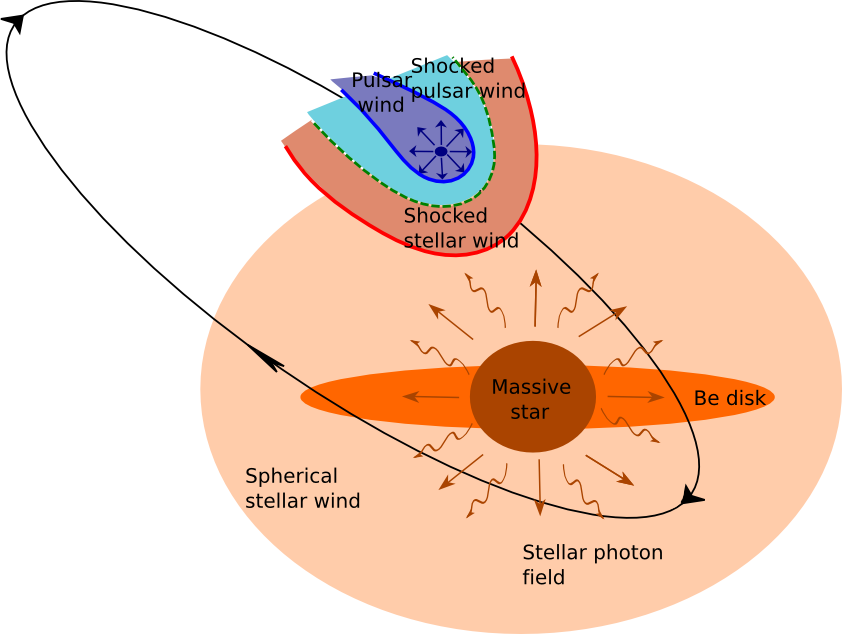}
\caption{Schematic view of a gamma-ray binary. The pulsar wind interacts with the stellar wind,  photon field and circumstellar disk if the companion is a Be star. \label{fig:gamma_binary}
}
\end{center}
\end{figure}

The last decade revealed a new group of gamma-ray emitters, composed of a fast-rotating pulsar and a massive star. The emission, which peaks in the MeV band, arises from the shocked region between the stellar wind and the pulsar wind. Particle acceleration at the relativistic shock creates emission up to several TeV, the highest emission observed for binary systems. Figure~\ref{fig:gamma_binary} presents a schematic view of interaction region.
A handful of these systems have been discovered so far: PSR B1259-63 \citep{2005A&A...442....1A}, LS 5039 \citep{2005Sci...309..746A} ,LSI +61$^o$ 303 \citep{2006Sci...312.1771A}, HESS J0632+057 \citep{2011ApJ...737L..11B}, 1FGL J1018.6-5856 \citep{2010ApJS..188..405A}, and more recently HESS J1832-093 \citep{2016MNRAS.457.1753E} and LMC P3 in the LMC \citep{2016ApJ...829..105C}. While pulsed emission has been observed for PSR B1259 (thanks to its wide orbit), and LSI+ 61$^o$ 303 has shown magnetar type flares \citep{0004-637X-744-2-106}, the nature of the compact object is not firmly established in the other systems. However, given the very similar emission patterns, the colliding wind scenario is now firmly established. As such, the major aspects of the colliding wind structure and the high-energy emission mechanisms have been assessed, and while many questions remain, the study of  gamma-ray binaries has matured enough to become a new window on pulsar wind physics.  These systems provide a unique opportunity to study otherwise very elusive pulsar winds. The binary interaction typically takes place around a fraction of  AU from the pulsar (or about $10^4$ times the light cylinder), about 5 orders of magnitude closer than for pulsars interacting with the ISM or supernova remnants.  Modeling strongly benefits from information provided by orbital variability and  the well constrained environment created by the companion star, both in terms of density and photons fields (at least compared to a typical region of the ISM or a supernova remnant).  While an excellent review can be found in \cite{2013A&ARv..21...64D}, this Section provides some updates and focuses on how to determine the pulsar wind properties from the emission of gamma-ray binaries.

\subsection {$\gamma$-ray binaries : puzzling observations}
\begin{figure}[htbp]
\begin{center}
\includegraphics[width=0.7\columnwidth]{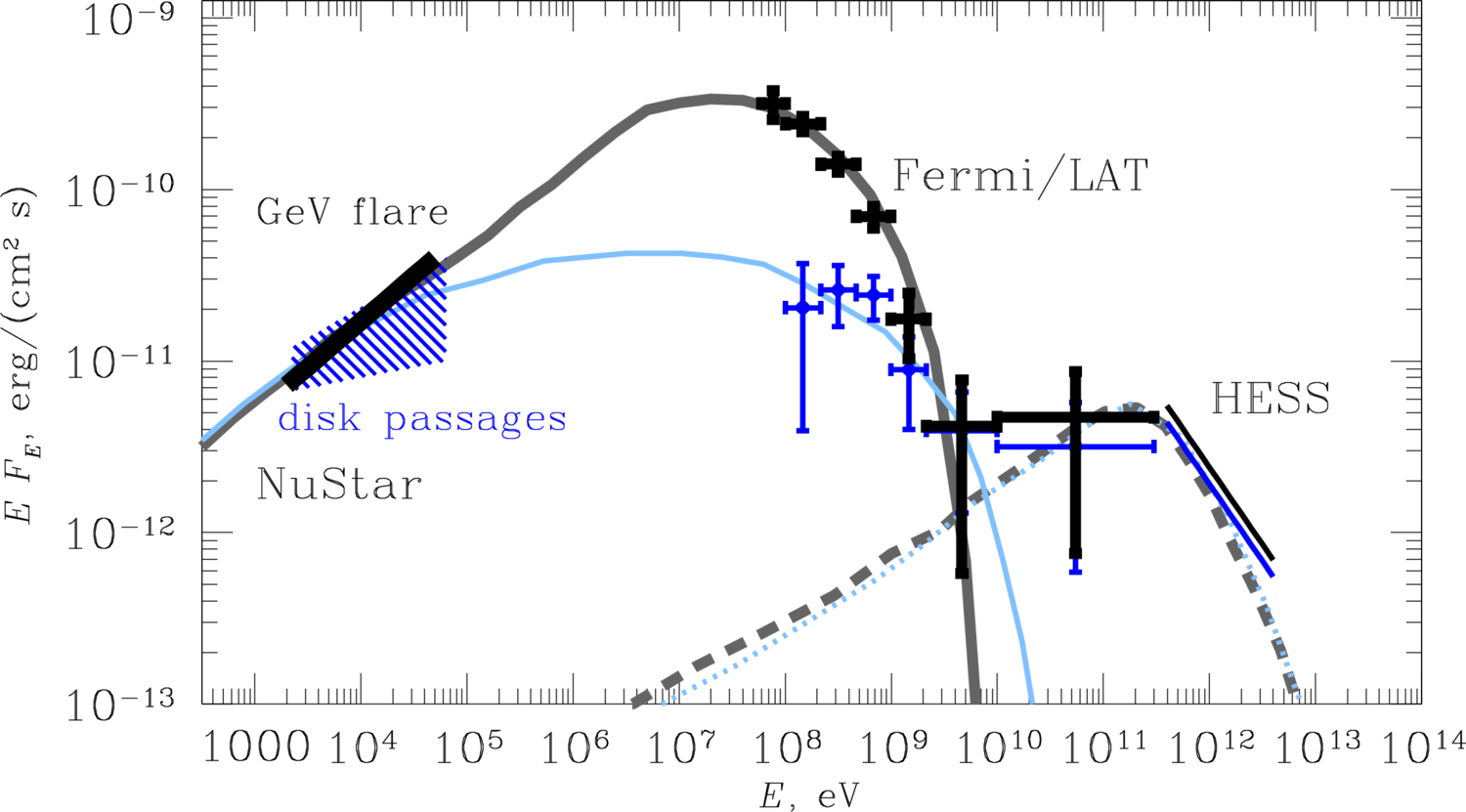}
\caption{High energy spectral energy distribution of PSR B1259-63, during the disk passages (blue) and flares (grey) (from \citet{2015MNRAS.454.1358C}). \label{fig:SED_PSR}
}
\end{center}
\end{figure}

While most of the energy is emitted in MeV photons (hence the name), gamma-ray binaries emit all the way from radio to a few TeV. Figure~\ref{fig:SED_PSR} shows the high energy spectral energy distribution of PSR B1259-63 with emission resulting from the electrons accelerated at the shocks between the winds. Acceleration at the relativistic shock produces synchrotron emission (up to a few GeV) and Inverse Compton emission on seed photons of the massive star (up to a few TeV).   

\begin{figure}[htbp]
\begin{center}
\includegraphics[width=0.7\columnwidth]{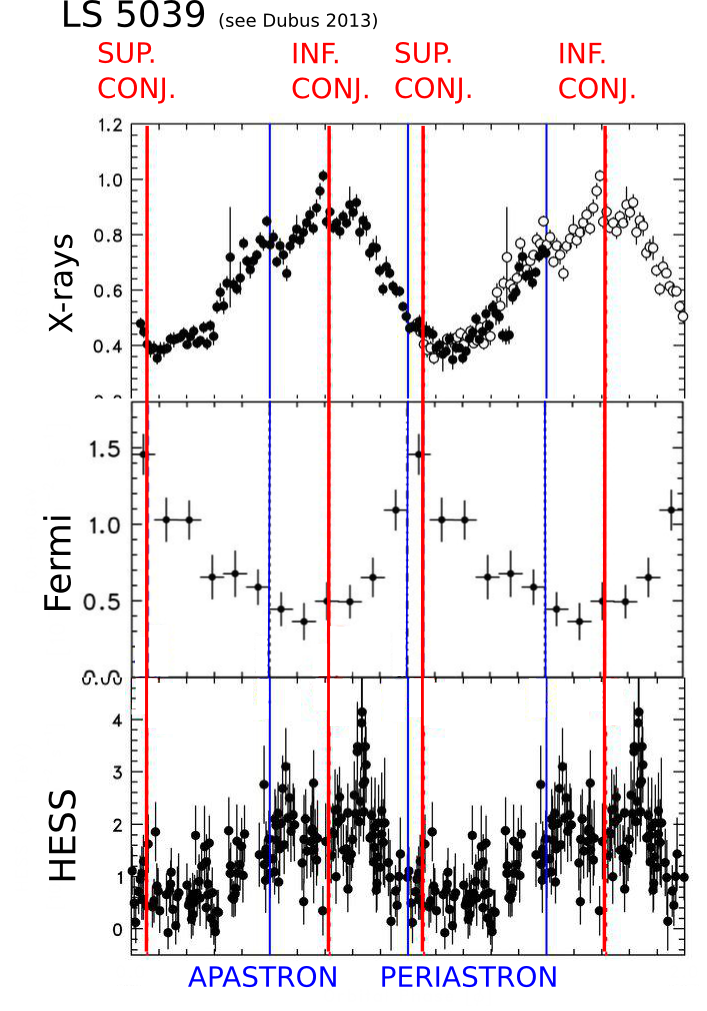}
\caption{X-ray, GeV and TeV modulation for LS 5039. (Adapted from \citet{2009ApJ...697L...1K,2009ApJ...706L..56A} and \citet{2006A&A...460..743A}). \label{fig:light_LS}
}
\end{center}
\end{figure}

In theses systems, most of the emission (with the exception of the low frequency radio band \citep{2015MNRAS.451...59M}) shows orbital variability. Figure~\ref{fig:light_LS} shows lightcurves of LS 5039, a binary with a 3.9 day period:  X-ray and TeV emission show similar orbital variability, in opposition with the variability observed with $Fermi/LAT$.  The MeV emission in LS 5039, with an exponential cutoff at a few GeV before the harder but fainter TeV emission cannot be reconciled while assuming emission from a single population \citep{2008A&A...477..691D}. The different origin of the GeV and TeV emission is also suggested by the absence of emission below 200 GeV  in HESS J0632+057  \citep{2016arXiv160108216M} and around periastron in PSR B1259-63, while both show strong TeV emission.  

LSI+61303 shows additional variability on superorbital timescales \citep{2012ApJ...747L..29C}, attributed to long-term variations in the Be disk surrounding the companion star \citep{2015A&A...575L...6P}. Interactions with the Be disk are likely responsible for the suborbital optical line variability in HESS J0632+057 \citep{2015ApJ...804L..32M}. 

In PSR B1259-63, which has a 3.4 year period, X-rays, TeV and radio flares are observed about 15 days before and after periastron. Optical spectroscopy indicates these flares are consistent with the disruption of the Be disk as the pulsar crosses it \citep{2016MNRAS.455.3674V}.  About 60 days after periastron, $Fermi/LAT$ detects strong emission with no counterpart at any other wavelength \citep{2011ApJ...736L..11A}.  This flare, which contains almost all the spindown power of the pulsar,  has been confirmed at the 2014 periastron passage \citep{2015ApJ...811...68C} with a similar flux level and spectral shape but slightly different temporal evolution.

Extended radio emission traces the contours of the shocked region \citep{2006smqw.confE..52D,2011ApJ...732L..10M}. Extended X-ray emission has also been found for most (if not all) of these systems but its origin remains unclear  \citep{2014AN....335..301K}.

\subsection{ Ingredients for successful models}

The wide variety of observational constraints is both a challenge and a tremendous opportunity to understand these systems.  Until recently, studies have either focussed on sophisticated one-zone emission models or aimed at reproducing the hydrodynamical  structure of the colliding wind region, with little connecion to the non-thermal emission. While both approaches have been fruitful, they have not been able to consistently reproduce the  orbital variability and spectral features from radio up to TeV.   The next sections describe both approaches and how they can be combined to yield a better understanding of pulsar wind physics.

The first ingredients are the non-thermal emission processes, dominated by leptonic processes \citep{2006A&A...456..801D}, causing a synchrotron bump and inverse Compton bump.  Models have included refined  representations of the ambient photon field, including self-Compton on the nebula  \citep{2010A&A...519A..81C} and the infrared photons of the Be disk when present  \citep{2012MNRAS.426.3135V}. Reproducing the TeV lightcurve of LS 5039 requires anisotropic inverse Compton emission  \citep{2008A&A...477..691D} and an extended emission region. The latter is necessary to prevent complete absorption by pair production at  superior conjunction in LS 5039 (when the compact object is behind the massive star).  Inverse-Compton pair cascades could extend the emission region \citep{2006MNRAS.371.1737B,2010A&A...519A..81C}. Or, the  emission could also result from one or more  distant emission regions \citep{2013A&A...551A..17Z}.  Similarly, the absence of occultation features in the X-ray lightcurves suggests an extended emission region \citep{2007A&A...473..545B,2011MNRAS.411..193S}.  As the emission originates from the relativistic shocked pulsar wind, Doppler boosting is also at work at inferior conjunction (when the pulsar wind points toward the observer) and explains some of the GeV modulations \citep{2010A&A...516A..18D}.

 Single zone models have so far failed to consistently reproduce the full variability of these systems, even when cascade emission is allowed. The exponential cutoff around a few GeV is hard to reconcile with the hard TeV emission. A single particle distribution with IC, synchrotron  and adiabatic  cooling fails to reproduce  this spectral feature \citep{2008MNRAS.383..467K,2013A&A...551A..17Z}.   The presence of an additional emission component has been suggested. While the GeV emission  is analogous to  pulsar magnetospheric emission observed by $Fermi$, its orbital variations are hard to explain \citep{2012ApJ...749...54H}.   The emission of the unshocked pulsar wind \citep{2007MNRAS.380..320K}   has been considered but is insufficient \citep{2008APh....30..239S}.  A shocked mono-energetic  component has also been considered \citep{2013A&A...557A.127D}.  An alternative path is the presence of a different acceleration region, with different properties in terms of magnetic field, photon field and velocity structure \citep{2013A&A...551A..17Z}.

Over the years, it has become clearer that the only path towards fully understanding the emission in gamma-ray binaries would require an elaborate model for the geometry of the system, coupled with a refined model for the non-thermal emission. 

The flow dynamics in gamma-ray binaries is dominated by the double shock structure. It  shares many similarities with colliding stellar winds, which have been studied for decades. Simulations of colliding wind binaries \citep{2009MNRAS.396.1743P,2011MNRAS.418.2618L} show strong instabilities which may yield to mixing in the winds \citep{2010MNRAS.403.1873Z} and affect the spiral structure expected at large scales \citep{2012A&A...546A..60L,2012A&A...544A..59B}.  

However, the relativistic nature of the pulsar wind affects the structure of the interaction region. Relativistic hydrodynamics induces more complex shock patterns, where parallel velocities (to the shock normal) play a role. Multidimensional simulations in the  ultrarelativistic regime of pulsar winds (Lorentz factor $\simeq 10^3-10^5$) are far beyond current computational and numerical capabilities. Some rescaling may be necessary even when focusing on the shocked winds,which have $\Gamma\lesssim 10$ close to the pulsar.  Simulations suggest a narrower opening angle for the pulsar wind \citep{2013A&A...560A..79L}, and a reacceleration of the pulsar wind up to its initial velocity \citep{2008MNRAS.387...63B}. Keeping in mind these intrinsic difficulties, relativistic simulations are crucial in order to determine Doppler boosting, which is a key ingredient to orbital modulation and maybe to explain the GeV flares in PSR B1259 \citep{2012ApJ...753..127K}. Simulations suggest the presence of a back shock behind the pulsar, which can provide an additional site for particle acceleration, further away from the binary. They also indicate the presence of a large scale spiral \citep{2012A&A...544A..59B,2016MNRAS.456L..64B}

\begin{figure}[htbp]
\begin{center}
\includegraphics[width=0.65\columnwidth]{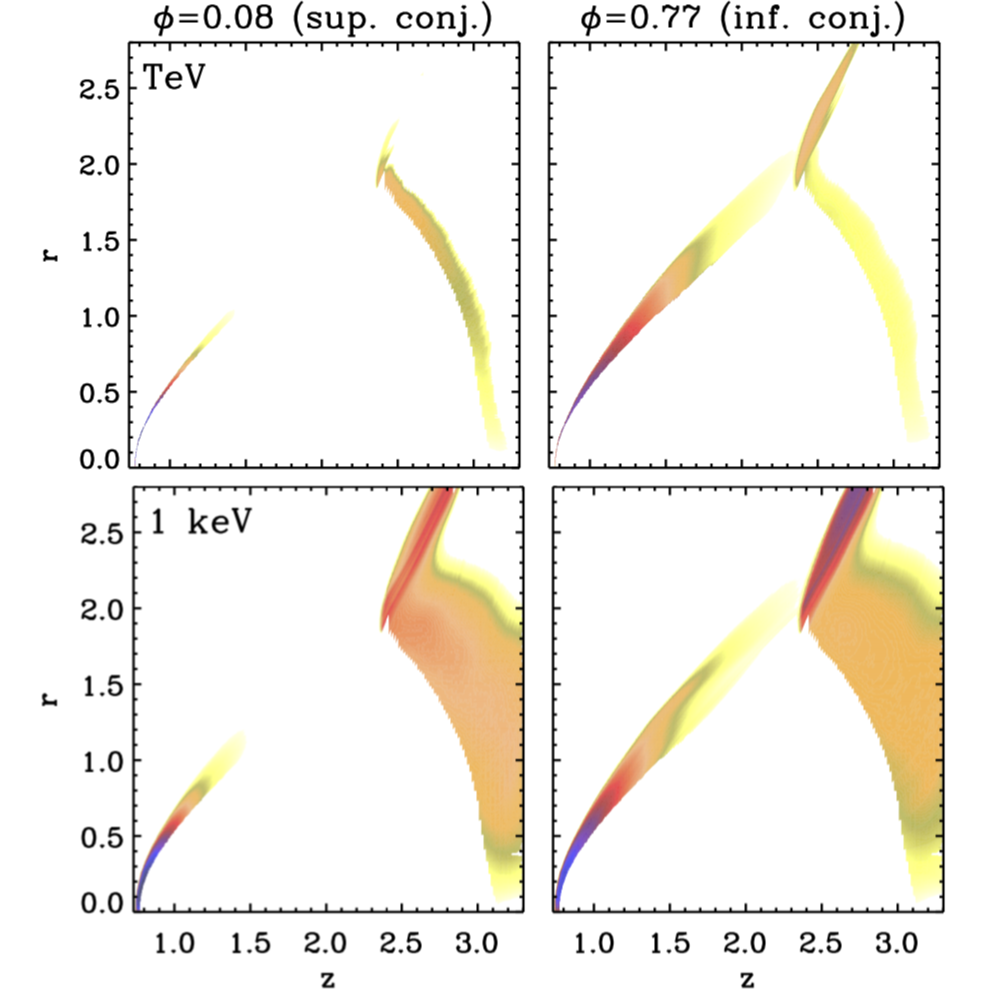}
\caption{TeV and X-ray emission maps for LS 5039 at superior and inferior conjunctions. \label{fig:simu_gamma1}. Image taken from \citet{2015A&A...581A..27D}.
}
\end{center}
\end{figure}

\begin{figure}[htbp]
\begin{center}
\includegraphics[width=0.9\columnwidth]{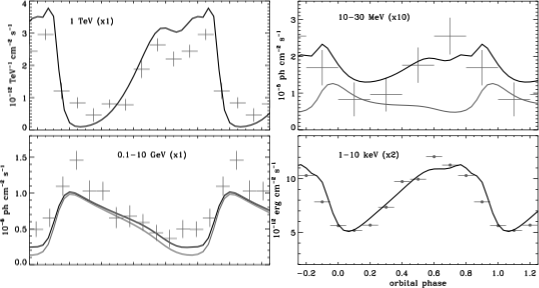}
\caption{Simulated lightcurves for LS 5039 with a powerlaw particle distribution (black) and an additional mono-energetic component (grey). Image taken from \cite{2015A&A...581A..27D}.\label{fig:simu_gamma}
}
\end{center}
\end{figure}

\cite{2015A&A...581A..27D} developed the first high energy emission model based on a fully three-dimensional relativistic simulation of LS 5039.  As described in Section \ref{sec:radiation}, the non-thermal emission is determined during post-processing, with a particle distribution injected at the relativistic shock and followed along the streamlines in the shocked pulsar wind. The model self-consistently represents adiabatic, inverse Compton and synchrotron cooling. The resulting emission takes into account pair creation, anisotropic inverse Compton and directly benefits from the simulation to model Doopler boosting.  Figs. ~\ref{fig:simu_gamma1} and \ref{fig:simu_gamma} show the resulting emission maps and lightcurves.  Comparison with observations suggested a strongly magnetized, rather slow pulsar wind ($\Gamma\simeq 10^3, \sigma \simeq 1$). While small discrepancies exist with observations, and radio emission will not be modeled properly without taking into account the magnetic field structure, the clear success of the combination of relativistic hydrodynamics with refined emission models shows the path forward in this field.


\section{Discussion}
\label{sec:discussion}

Over the past two decades, modelling of PWN has lead to a good understanding of the plasma flow in these systems. Predominantly, most studies have focused on the Crab nebula.  The global plasma flow and dynamics in this system are well described by the simulations. The origin of its jet, the torus, the inner knot and the wisps are understood. However, other features as the knots of the inner ring or the ``thin wisp'' \citep{hester-95} are not present in the simulations. Also, the morphology of other PWN appears to be more complex. In the Vela PWN for instance, the receeding jet appears to be brighter than the one pointed to the observer \citep{Durant_2013}, at odds with the expectation of Doppler beaming in plasma flow models.

Recently, an important step forward was done by simulating the Crab nebula in 3D. The 3D randomisation of the magnetic field leads to magnetic dissipation inside the PWN. This removes the constraint of low magnetization of the pulsar wind imposed by 1D and 2D axisymmetric models and is likely the solution to the longstanding ``$\sigma$-Problem''. As always, a deeper understanding comes with many new questions. Also, several older questions remain unanswered. We will summarize the ones we consider most important in the following: 

\begin{enumerate}

\item \textbf{Where and how does non-thermal particle acceleration take place? How does the particle spectrum evolve in the nebula, particularly along the termination shock? }

Magnetic reconnection appears to be the main process by which particles are accelerated in the nebula. It remains unclear where and how this happens. In particular it is unclear if different acceleration mechanisms take place at different latitudes along the termination shock and what the contribution of second order Fermi acceleration within the body of the nebula is.

\item \textbf{What is the effect of the feedback of the acceleration on the plasma flow?}

As a significant fraction of the total energy goes into accelerating particles, their back reaction on the plasma could be important. To date, only test particle simulations have been performed, neglecting the feedback of the accelerated particles on the global plasma flow. The simple ideal MHD approximations will need to be extended to a more realistic plasma description.

\item \textbf{Where within the nebulae do the gamma-ray flares originate? Are these exceptional events with little impact on the overall evolution/emission of PWN or a high-energy tail of the main dissipative processes?  }

Significant progress has been made in understanding the processes that can lead to explosive gamma-ray flares. However, if and where within the nebula these processes happen remains unclear. It is puzzling that no gamma-ray flares have been discovered from any other isolated PWN than the Crab nebula. It is not clear if similar processes are responsible for Crab nebula flares and the GeV flares of the binary PSR B1259-63.

\item \textbf{What is the origin of radio electrons? Are they supplied by the pulsar wind and if not then why does the spectrum of emission not show a discontinuity between the radio and optical parts? }

The hard energy spectrum of the radio emitting electrons suggests that they are accelerated via magnetic reconnection. However, where these electrons come from remains an open question. If the electrons are provided by the wind their large number is in conflict with magnetospheric models. An alternative is that electrons are stripped from the gas and dust in the nebula's filaments. In this case, however, spectral continuity with the higher energy particle population would be a mystery, and just as mysterious would be the similarity between the radio spectra of PWN and those of bow-shock PWN, where there are no filaments from which low energy particles could come.

\item \textbf{What is the structure of termination shock in the polar region? }

In the polar region the pulsar wind has no stripes and hence a very high magnetization is expected both upstream and downstream of the termination shock. It is unclear what the fate of the highly magnetized flow injected into the nebula in this region is and if its magnetic energy can be dissipated on a light-crossing time.

\item \textbf{How many electrons/positrons escape into the ISM, and with which energies? Can PWN potentially explain the positron excess?}

The interaction of PWN with the surrounding medium is complex, as they evolve within the remnant of their progenitors explosions. It is unclear how many electrons can escape the magnetic confinement of these systems and if their number is sufficient to explain the positron excess observed at Earth. Of course, if the radio emitting particles, which are the relevant ones for this matter, come from filaments, and are therefore all electrons (no positrons), then PWN cannot contribute to the positron excess. 

\item \textbf{Is an energetically significant amount of non-thermal ions present in PWN?}

To date, no observational signature of ions has been found in PWN. However, due to their low radiative efficiency, there could still be an energetically significant number of ions present in the pulsar wind and it is interesting to notice that these would be ions with energies in the PeV range.  

\end{enumerate}

The search for answers of these questions is ongoing. Fortunately, deeper observations and increasingly realistic simulations are powerful tools at our disposal in this quest.

\section*{Acknowledgments}
OP is supported by the ERC synergy grant ``BlackHoleCam: imaging the Event Horizon of Black Holes'' (Grant No. 610058).
BO acknowledges support from the INFN -- TEONGRAV initiative (local PI: Luca Del Zanna) and from the University of Florence grant ``Fisica dei plasmi relativistici: toeria e applicazioni moderne''.  
Support for AL was provided by an Alfred P. Sloan Research Fellowship, NASA ATP Grant NNX14AH35G and NSF Collaborative Research Grant 1411920 and CAREER grant 1455342.  

\bibliographystyle{mnras}
\bibliography{bibliography/converted-to-latex.bib}

\end{document}